\documentclass[fleqn,usenatbib]{mnras}

\usepackage{newtxtext,newtxmath}
\usepackage[T1]{fontenc}
\usepackage{ae,aecompl}

\usepackage{graphicx}	% Including figure files
\usepackage{amsmath}	% Advanced maths commands
\usepackage{amssymb}	% Extra maths symbols
\usepackage{ascmac}
\usepackage{siunitx,physics}
\usepackage{xcolor}
\usepackage{ulem}

\newcommand{\Rc}{R_\text{solid}}
\newcommand{\Mc}{M_\text{solid}}
\newcommand{\Msolid}{M_\text{solid}}
\newcommand{\Mp}{M_\text{p}}

\newcommand{\Matm}{M_\text{atm}}
\newcommand{\dMesc}{\dot{M}_\text{esc}}
\newcommand{\Lxuv}{L_\text{XUV}}
\newcommand{\rhos}{\rho_\text{solid}}

\newcommand{\RB}{R_\text{B}}
\newcommand{\RH}{R_\text{H}}

\newcommand{\rhoout}{\rho_\text{out}}
\newcommand{\Tout}{T_\text{out}}
\newcommand{\NBLad}{\nabla_\text{ad}}
\newcommand{\NBLrad}{\nabla_\text{rad}}

\newcommand{\rsat}{r_\text{sat}}

\newcommand{\rl}{r_l}
\newcommand{\cn}{c_\text{n}}
\newcommand{\cv}{c_\text{v}}
\newcommand{\cl}{c_l}
\newcommand{\Rn}{\mathcal{R}_\text{n}}
\newcommand{\Rv}{\mathcal{R}_\text{v}}
\newcommand{\Rg}{\mathcal{R}_\text{g}}
\newcommand{\mun}{\mu_\text{n}}
\newcommand{\muv}{\mu_\text{v}}
\newcommand{\psat}{p_\text{sat}}
\newcommand{\pn}{p_\text{n}}

\newcommand{\rhogas}{\rho_\text{gas}}
\newcommand{\rhotot}{\rho_\text{tot}}

\newcommand{\mH}{m_\text{H}}
\newcommand{\HHO}{\text{H}_2\text{O}}

\newcommand{\HH}{\text{H}_2}
\newcommand{\OO}{\text{O}_2}
\newcommand{\He}{\text{He}}
\newcommand{\XH}{X_\text{H}}
\newcommand{\XHe}{X_\text{He}}
\newcommand{\XO}{X_\text{O}}

\newcommand{\XHHO}{X_{\HHO}^0}

\newcommand{\XHH}{X_{\HH}}

\graphicspath{{Figs/}}
\title[Aqua planet formation with nebular-origin water]{Formation of aqua planets with water of nebular origin: Effects of water enrichment on the structure and mass of captured atmospheres of terrestrial planets
}
\author[]{
Tadahiro Kimura,$^{1}$\thanks{t.kimura@eps.s.u-tokyo.ac.jp}
Masahiro Ikoma,$^{1,2}$\thanks{ikoma@eps.s.u-tokyo.ac.jp}
\\
$^{1}$
Department of Earth and Planetary Science, Graduate School of Science, The University of Tokyo, 7-3-1 Hongo, Bunkyo-ku, Tokyo 113-0033, Japan
\\
$^{2}$
Research Center for the Early Universe (RESCEU), Graduate School of Science, The University of Tokyo, 7-3-1 Hongo, Bunkyo-ku, Tokyo 113-0033, Japan
}

\date{Accepted XXX. Received YYY; in original form ZZZ}

\pubyear{2020}

\begin{document}
\label{firstpage}
\pagerange{\pageref{firstpage}--\pageref{lastpage}}
\maketitle

% Abstract of the paper
\begin{abstract}
% Background & Context
%\kimura{
Recent detection of exoplanets with Earth-like insolation attracts growing interest in how common Earth-like aqua planets are beyond the solar system.
%Because of recent detection of exoplanets with Earth-like insolation, there is a growing interest in how common Earth-like aqua planets are beyond the solar system.
% Motivation & Goal
While % theoretical predictions so far \kimura{assume} that 
terrestrial planets are often assumed to capture icy or water-rich planetesimals, 
a primordial atmosphere of nebular origin itself %is capable of producing
can produce water through oxidation of the atmospheric hydrogen with oxidising %rocky materials 
minerals from incoming planetesimals or the magma ocean.
% Current status (Previous studies)
Thermodynamically, normal oxygen buffers produce water comparable in %mass 
mole number to or more than hydrogen.
%Given this process, 
Thus, the primordial atmosphere 
would likely be highly enriched with water vapour; 
however, the primordial atmospheres %of terrestrial planets 
have been always assumed to have the solar abundances.
% Aim & Method
Here we integrate the 1D structure of such an enriched atmosphere %connected to the surrounding nebula gas 
of sub-Earths embedded in a protoplanetary disc around an M dwarf of 0.3$M_\odot$
and investigate the effects of water enrichment on the atmospheric properties with focus on %the 
water amount.
%amount of water.
%\kimura{In this study 
%We consider especially sub-Earth-mass planets around 
%\ikoma{an M dwarf of} 0.3$M_\odot$.}%solar-mass star.}
% Results
We %have found 
find that the %mass of the \kimura{80\%} H$_2$O 
well-mixed, highly-enriched atmosphere is %larger 
more massive
by a few orders of magnitude than %that for 
the solar-abundance atmosphere, and that even a Mars-mass
%\remove{sub-Earth-mass}
planet can obtain water comparable %in mass 
to the present Earth's oceans.
% \remove{Moreover, even after considering the atmospheric mass reduction process due to disc gas dissipation and stellar irradiation, sub-Earth and even sub-Mars planets can keep the sufficient amount of atmosphere and water.
% Conclusions
Although close-in Mars-mass planets likely lose the captured water via disc dispersal and photo-evaporation,
these results suggest 
%the possibility that there may exist many 
that there are more %terrestrial exoplanets 
sub-Earths with Earth-like water contents than previously predicted.
How much water terrestrial planets really obtain %by this process 
and retain against subsequent loss, however, depends on 
efficiencies of water production, %efficiency, %and 
%material 
mixing in the atmosphere and magma ocean, 
and photo-evaporation, detailed investigation for which
%of which detailed investigation 
should be made in the future.
%}
%253 words...
\end{abstract}

% Select between one and six entries from the list of approved keywords.
% Don't make up new ones.

%%%%%%%%%%%%%%%%%%%%%%%%%%%%%%%%%%%%%%%%%%%%%%%%%%

%%%%%%%%%%%%%%%%% BODY OF PAPER %%%%%%%%%%%%%%%%%%

\section{Introduction} \label{sec:intro}
Planetary climate depends greatly on the amount of ocean water.
The present Earth's oceans account for only 0.023~\% of the planetary mass.
Such a small amount of ocean water allows continents to appear on the Earth.
Continental weathering plays a crucial role in the geochemical carbon cycle and thereby keeps the Earth's climate stable over a geological timescale~\citep{Walker+1981,Caldeira1995}. 
If the Earth had oceans three times more massive 
than the present, all the continents would be submerged in the global ocean \citep[e.g.,][]{Maruyama+2013}.
Recent theories predict that terrestrial planets covered completely with oceans have extremely hot or cold climates 
\citep{Abbot+2012,Kaltenegger+2013,Alibert2014,Nakayama+2019}.
Thus, stable, temperate climates are possible in a relatively narrow range of ocean mass.

A widespread idea is that water is brought from afar to terrestrial planets in the habitable zone (HZ). 
This is partly because it is difficult for planets in HZ, which is interior to the snowline, to obtain water in situ.
As for planetary systems that have giant planets exterior to the snowline like the solar system, the giant planets scatter icy or water-rich planetesimals gravitationally to deliver water to HZ.
Indeed, direct $N$-body simulations for planetesimal accretion under the gravity of a Jupiter-mass planet at 5.2~AU \citep[e.g.][]{Raymond+2004,Zain+2018} demonstrate that rocky planets in HZ, in most cases, obtain more water by a factor of $\sim$ 3 to 100 than the Earth's oceans.
This suggests that terrestrial planets having oceans similar or more in amount than the Earth could be common around Sun-like stars.

In contrast, 
%\kimura{
exoplanet surveys show that the occurrence rate of giant planets around M dwarfs is lower than those around Sun-like stars \citep{Endl+2006,Cumming+2008,Mulders+2015}.
This is consistent with the theoretical prediction from core accretion models that massive enough cores for runaway gas accretion are rarely formed in less massive circumstellar discs around low-mass stars \citep[e.g.,][]{Ida+2005}. 
Therefore, the process of water delivery by giant planets described above is considered not to commonly occur around M dwarfs. 
%\ikoma{Also, there is no barrier against migration of icy planetesimals and embryos which giant planets would act as.}
As a result, 
%the low-mass circumstellar discs produce 
%extremely dry rocky planets \ikoma{form in situ or water-dominated planets form via migration} 
%and also bring icy materials 
%from beyond the snowline. 
%by orbital migration to form water-dominated ($>$~50~\%) planets.
%If this is the case, 
rocky planets with Earth-like water contents are rare in the habitable zones around M dwarfs \citep[e.g.,][]{Tian+Ida2015}.
%}

Other than the planetesimal origin of water,
the primordial atmosphere that a protoplanet captures from the circumstellar disc \citep{Sasaki1990,Ikoma+Genda2006} 
is capable of producing water.
The atmospheric hydrogen is oxidised to produce water by oxides in vaporising materials from planetesimals passing through the atmosphere and those in the magma ocean covering the proto-planetary surface.
While being dependent on the kind of oxide, the equilibrium partial pressure ratio $P_{\HHO}/P_{\HH}$ is on the order of unity for normal iron oxides found in meteorites and the Earth's crust.
This means that rocky planets can acquire water in situ even inside the snowline, provided they are embedded in a circumstellar disc. 
The water thus produced is called the captured water hereafter in this study.

To see how much water a rocky protoplanet captures, we investigate the structure of the primordial atmosphere enriched with water in this study, 
since the atmosphere's structure controls its mass for an atmosphere equilibrated with a circumstellar gaseous disc. 
As for the solar-abundance (or unenriched) primordial atmosphere, detailed investigation was conducted previously \citep[][]{Hayashi+1979,Nakazawa+1985,Ikoma+Genda2006}
\citep[also see the recent reviews,][]{Massol+2016,Ikoma+2018}.
The most important finding is that as long as the atmosphere is optically thick enough, the atmospheric mass is closely related to the thermal state of the atmosphere, which is controlled by the opacity and the energy flux; the latter is supplied predominantly by incoming planetesimals in accretion stages. Another finding is low sensitivity to the outer boundary conditions of the atmosphere (or the disc gas conditions). 
Numerical models of the 1D atmospheric structure show that protoplanets of $\gtrsim$~0.3~$M_\oplus$ have such thick atmospheres in the normal ranges of the opacity (grain depletion factor $f$ of 0.01--1; see Eq.~\eqref{eq:kappa} for definition) and planetesimal accretion rates ($10^{20}$--$10^{26}$~erg/s in terms of luminosity)~\citep{Ikoma+Genda2006}.

The atmospheric properties for small protoplanets ($< 0.3 M_\oplus$) are qualitatively different from the above
~\citep{Ikoma+Genda2006}. 
The atmosphere is nearly isothermal and its mass is smaller by approximately 3--5 orders of magnitudes than in the case of an Earth-mass protoplanet.
Also the atmospheric structure and mass are sensitive to the nebular gas density and, thus, such a small-mass protoplanet loses most of the atmosphere as the nebular gas decreases.
Therefore it is predicted that protoplanets with masses less than a few Mars masses are unable to have massive primordial atmospheres.

Those previous studies, however, considered only  atmospheres with the solar element abundances and ignored the effects of water vapour enrichment on the structure and mass of the atmosphere.
This was previously investigated in the context of gas giant formation~\citep{Hori+Ikoma2011,Venturini+2015,Chambers2017}.
They showed that the mass of the proto-gas giant envelope significantly increases by the pollution due to the effects of the increase in mean molecular weight and reduction in heat capacity of the envelope gas. 
The effect of enhanced opacity was found to be negligible because the envelope is almost entirely convective.
Consequently, the critical core mass for runaway gas accretion decreases by one to two orders of magnitude in the Jupiter-forming region if the envelope is so polluted that the envelope's metallicity $Z$ exceeds $\sim 0.6$.

Likewise, the water production through oxidation of hydrogen in the primordial atmosphere of terrestrial planets is expected to be effective in increasing the atmospheric mass and, thus, water mass significantly.
This study is aimed at quantifying that effect, 
focusing on planets of $\sim 0.1$--$1.0$~$M_\oplus$, which are expected to be abundant around M dwarfs~\citep{Raymond+2007,Ida+2005,Alibert+Benz2017,Miguel+2019}.
The rest of this paper is organised as follows.
In Section~\ref{Sec:Method}, we describe the numerical model of the atmospheric structure.
In Section~\ref{Sec:Results}, we show the results of our calculations especially on the effect of water vapour enrichment on the atmospheric mass and structure and also estimate the water amount in the atmosphere for various planetary masses and boundary conditions.
In Section~\ref{Sec:Discussioin}, we discuss the process to enrich the atmosphere with water especially for planets inside the snowline, and also the importance of some processes ignored in this study.
Finally, we conclude this study in Section~\ref{Sec:Conclusioin}.

\section{Method} \label{Sec:Method}
We consider a protoplanet with a spherically symmetric structure that is embedded in a protoplanetary disc and subject to continuous planetesimal bombardment. 
The protoplanet has an atmosphere on top of a rigid body with a density $\rhos$ of 3.2~$\si{g/cm^3}$,
assuming that the protoplanet is undifferentiated like previous studies~\citep[e.g.,][]{Hori+Ikoma2011,Venturini+2015}. 
Note that the atmospheric structure and mass are insensitive to choice of $\rhos$.
The atmosphere is in hydrostatic and thermal equilibria; 
its energy source is assumed to be the kinetic energy of incoming planetesimals, which is released at the bottom of the atmosphere. 
Other energy sources including cooling of and radioactive decay in the rocky body are known to be negligible in the main accretion stage \citep[][]{Ikoma+Genda2006,Ikoma+Hori2012}.
For atmospheric masses of $\lesssim$ 20-30~\% of the solid-body mass, the assumptions both of hydrostatic and thermal equilibria are valid, as verified by previous studies \citep[e.g.,][]{Ikoma+2000}. This means that
%the atmospheric properties at this stage does not depend on its history as suggested by previous studies~\citep[e.g.][]{Ikoma+2000}. 
the properties of less massive atmospheres of interest in this study 
%do not depend on
are insensitive to the evolution history.
Thus we do not solve the time evolution and instead, 
%just solve the static structure of the atmosphere using the 
integrate equations without time derivatives (i.e., entropy change) as described below.
We also ignore the effects of planetary migration.
The atmosphere is uniform in element abundance and composed of H, He, and O from the disc gas and O from the planetesimals and the magma ocean below the atmosphere. 
In this study, we ignore the effect of ingassing to the magma ocean, for simplicity (see Sect.~\ref{Sec:Discussioin}).

\subsection{Basic Equations}%%%%%%%%%%%%%%%%%%%%%%%%%%%%%%%%%%%%%%%%
The structure of the atmosphere is calculated with the following equations;
\begin{align}
 \dv{P}{r}   &= -\frac{GM_r \rho}{r^2}, \label{eq:hydrostat} \\
 \dv{M_r}{r} &= 4\pi r^2 \rho,\label{eq:mass-conserve} \\
 \dv{T}{r}   &= -\frac{GM_r \rho}{r^2}\frac{T}{P}\nabla \label{eq:dTdr},
\end{align}
where $r$ is the radial distance from the protoplanet centre,
$M_r$ is the mass inside a sphere of radius $r$, 
$G$ is the gravitational constant $(=\SI{6.67e-8}{cm^3/g.s^2})$,
$P$, $T$, and $\rho$ are the pressure, temperature, and density of the atmospheric gas, respectively,
and $\nabla=\dv*{\log T}{\log P}$.
Since we consider vapour condensation, we use either of two types of density, namely, the density only of the gaseous components, $\rhogas$, 
or the total density including condensed water, $\rhotot$.
When using the former, we assume that the condensates (or water/ice drops) precipitate quickly and thus only the gaseous components contribute to pressure. 
When using the latter, we assume that the condensates remain in the atmosphere; thus, both in Eqs.~\eqref{eq:hydrostat} and \eqref{eq:mass-conserve}, $\rhotot$ should be used.
We show this effect on the atmospheric mass in Sec.~\ref{sec:pseudo}.

The temperature gradient $\nabla$ is the smaller of adiabatic temperature gradient $\NBLad$ and radiative temperature gradient $\NBLrad$, which are expressed, respectively, by
\begin{align}
 \NBLad  &= \qty( \pdv{\ln T}{\ln P})_S, \\
 \NBLrad &= \frac{3\kappa L P}{64\pi \sigma T^4 G M_r},
\end{align}
where $L$ is the total energy flux passing through the spherical surface of radius $r$ (simply the energy flux, hereafter), $\kappa$ is the Rosseland-mean opacity, and $\sigma$ is the Stefan-Boltzmann constant $(= \SI{5.67e-11}{erg/(cm^2.K^4.s)})$.
%\kimura{
%$L$ is assumed to be supplied only by the accreting planetesimals, and treated as an input parameter.
%Other energy sources, such as cooling of rocky bodies and radioactive decay would be negligible at this stage~\citep{Ikoma+Genda2006,Ikoma+Hori2012,Guillot+1995}.
%}

Water vapour condensation occurs if the partial pressure of water vapour 
exceeds the saturation vapour pressure (or the Clausius-Clapeyron relation) expressed by~\citep[e.g.,][]{Kasting+1984,Kasting1988}
\begin{equation}
 \psat = p_0 \exp( -\frac{\ell}{\Rv T}),
 \label{eq:psat}
\end{equation}
where $\ell$ is the specific latent heat of condensation, 
$\Rv = \Rg/\muv$ with gas constant $\Rg (= \SI{8.31e7}{erg/(mol. K)})$ and 
mean molecular weight of vapour $\muv$, and $p_0$ is a constant.
In this region, the temperature gradient can be written 
by the moist adiabat~\citep[e.g.,][]{Kasting1988} 
\begin{equation}
 \dv{\ln p}{\ln T}
  = \frac{1}{1 + \rsat/\epsilon}
    \frac{\cn + \qty{\cv + \frac{\ell^2}{\Rn T^2}
          (\rsat + \epsilon)}\rsat + \rl \cl}
         {\Rn + \frac{\ell}{T}\rsat}, \label{eq:moist-adiabat}
\end{equation}
where $\cn$, $\cv$ and $\cl$ are the specific heats of non-condensable molecules, condensable molecules (i.e., vapour), and condensates, respectively, $\Rn = \Rg/\mun$ with mean molecular weight of non-condensable components $\mun$.
$\rl$ is the mass ratio of condensates to non-condensable molecules per unit volume.
$\rsat$ and $\epsilon$ are defined as
\[
 \rsat = \frac{\Rn}{\Rv}\frac{\psat}{\pn} =: \epsilon \frac{\psat}{\pn},
\]
where $\pn$ is the partial pressure of non-condensable components.
The constant values needed for calculation are summarised 
in Table~\ref{tab:constants}.
Note that $p_0$ is calculated so that Eq.~\eqref{eq:psat} satisfies the triple point of water.
$\cn$ and $\cv$ are obtained by the chemical equilibrium calculation described in Section~\ref{sec:eos}.
When the condensates are assumed to precipitate immediately, $\rl = 0$ in Eq.~\eqref{eq:moist-adiabat}.

\begin{table}
    \caption{Thermodynamical constant values~\citep{Bohren+Albrecht1998}}   
    \label{tab:constants}
    \centering
        \begin{tabular}{c|c|c}\hline
             & $T < 273~\si{K}$ & $T > 273~\si{K}$ \\ \hline
            $p_0~[\si{erg/cm^3}$] &  $\num{3.48e13}$ & $\num{2.47e12}$ \\
            $\ell~[\si{erg/g}]$ & $\num{2.83e10}$ & $\num{2.50e10}$ \\
            $\cl~[\si{erg/(g.K)}]$ & $\num{2.09e7}$ & $\num{4.20e7}$ \\
            \hline
        \end{tabular}
\end{table}

\subsection{Equation of state} %%%%%%%%%%%%%%%%%%%%%%%%%%%%%%%%%%%%
\label{sec:eos}
We assume that the atmosphere consists of the three elements H, He, and O, and consider the following nine species, H, He, O, $\HH$, $\OO$, $\HHO$, $\text{H}^{+}$, $\text{O}^{-}$, $\text{e}^{-}$.
We calculate the chemical equilibrium values of thermodynamic quantities, assuming the atmospheric gas is a mixture of ideal gases, namely
\begin{equation}
    P = \frac{\rhogas kT}{\mu \mH}, \label{eq:eos}
\end{equation}
where $\mu$ is the mean molecular weight, $k$ is the Boltzman constant $(=\SI{1.38e-16}{erg/K})$, and $\mH$ is the proton mass (=$\SI{1.66e-24}{g}$).
For the calculation, we use the numerical code developed by \citet{Hori+Ikoma2011}.

\subsection{Opacity}%%%%%%%%%%%%%%%%%%%%%%%%%%%%%%%%%%%%%%%%%%%%%%%
As the opacity sources, we consider radiative extinction by gaseous molecules and dust grains floating in the atmosphere.
The gas opacity is the Rosseland mean opacity calculated from the absorption cross-section of each molecule.
We consider the line absorption of $\HHO$~\citep{Polyansky+2018} and the collision-induced absorption (CIA) of $\HH$-$\HH$ and $\HH$-$\He$~\citep{Karman+2019}.
For the calculation of Rosseland mean opacities, we use the open-source code ExoCross~\citep{Yurchenko+2018b}.
We assume that the dust grains consist of water ice, organics and minerals and adopt the opacity model for dust grains in protoplanetary discs developed by \cite{Semenov+2003}.
Given that the number and size of dust grains change due to dust growth and sedimentation in the atmosphere, 
we introduce a factor $f$ and express the total opacity $\kappa$ as
\begin{equation}
    \kappa = \kappa_\mathrm{gas} + f \kappa_\mathrm{grain},
    \label{eq:kappa}
\end{equation}
where $\kappa_\mathrm{gas}$ and $\kappa_\mathrm{grain}$ are the opacities of gas and grains, respectively.

In contrast to the previous studies
\citep[][]{Ikoma+Genda2006,Hori+Ikoma2011,Venturini+2015} 
which assume a constant value of $f$ through the atmosphere, we calculate the value of $f$ at each altitude,
following the method presented in \cite{Ormel2014}.
Here we briefly summarise the Ormel's model: 
Dust grains are released to the atmosphere by planetesimals and then settle down and grow via mutual collision. 
Thus, the mass and size distributions of the grains are determined by a balance between such source and sink in a steady state.
Adopting a single-size approximation such that the mass distribution of dust grains is characterised by a characteristic mass $m^\ast$, we obtain the differential equation for the radial distribution of $m^*$ as
\begin{equation}
    \pdv{m^*}{r} = -\frac{m^*}{v_{\text{settl}} T_{\text{grow}} }
        + \frac{m_{\text{dep}} - m^*}{\dot{M}_{\text{dep}} }
           \pdv{\dot{M}_{\text{dep}} }{r}, \label{eq:grain-mass}
\end{equation}
where $v_{\text{settl}}$ and $T_{\text{grow}}$ are, respectively, the settling velocity and growth timescale of dust grains, $m_{\text{dep}}$ is the mass of the smallest grains (or monomers), and $\dot{M}_{\text{dep}}(r)$ is the cumulative mass flux of the dust grains that have been released at altitude $r$.
The first and second terms on the right-hand side represent the growth and deposition of dust grains, respectively.
Such a single-mass approximation was verified to yield similar results to those obtained with detailed multi-size calculations \citep[e.g.,][]{Kawashima+Ikoma2018}.
Further details of this model are described in Appendix~\ref{sec:appendix}.

\subsection{Atmospheric Composition}%%%%%%%%%%%%%%%%%%%%%%%%%
As mentioned at the beginning of this section, the atmosphere is assumed to be uniform in element abundance and consist of the three elements H, He, and O; thus, the mass fraction of element $i$ being denoted by $X_i$,
\begin{equation}
    \XH + \XHe + \XO = 1.
\end{equation}
In this study, 
we use the water mass fraction $\XHHO$, instead of $\XO$, as an input parameter. 
Although the actual water mass fraction varies in the atmosphere, $\XHHO$ is defined by
\begin{equation}
    \XHHO = \frac{9}{8} \XO
\end{equation}
and thus constant through the atmosphere. We assume that only the abundance of O is non-solar, while the He/H ratio is solar \citep[= 0.385 $\equiv A$;][]{Lodders+2009}.
From the above two relations, $\XH$ is derived as
\begin{equation}
 \XH = \frac{1-8\XHHO/9}{1 + A}.\label{eq:XH}
\end{equation}

\subsection{Boundary Conditions}%%%%%%%%%%%%%%%%%%%%%%%%%%%%%%%%

The bottom of the atmosphere corresponds to the surface of the solid body whose radius $\Rc$ is calculated as
\begin{equation}
    \Rc \equiv \left( \frac{3 \Msolid}{4 \pi \rhos} \right)^{1/3}.
\end{equation}
The inner boundary condition for the atmospheric structure is given there by
\begin{equation}
    M_r = \Msolid 
    \hspace{3ex} \mbox{at} 
    \hspace{3ex} r = \Rc .
\end{equation}

The outer boundary radius is set to be the smaller of the Bondi radius $\RB$ and the Hill radius $\RH$, which are defined, respectively, as
\begin{align}
    \RB &= \frac{G\Mp}{k\Tout/(\mu\mH)}, 
    \label{eq:RB} \\
    \RH &= \qty(\frac{\Mp}{3M_*})^{1/3} a,
\end{align}
where $\Mp$ is the protoplanetary total mass, $\Tout$ is the temperature at the boundary,
%\remove{$\mu$ is set to $2.34$ in all calculations,} 
$M_\ast$ is the central star's mass, and $a$ is the semi-major axis of the protoplanet.
Given the mass of the atmosphere is smaller than that of the solid part of the protoplanet (called the solid protoplanet, hereafter), we use the solid protoplanet mass $\Mc$, instead of $\Mp$, for the calculation of $\RB$ and $\RH$. This hardly affects results in this study. 
In the simulations below, $\Mp$ is so small that the outer boundary radius is always equal to $\RB$ ($< \RH$). 
We set $\mu$ to the disc value (i.e., 2.34), assuming
%\remove{Since we assume} 
that the 
%\remove{primordial} 
atmosphere is smoothly connected to the surrounding disc gas. 
%\remove{, the mean molecular weight $\mu$ in Eq.~\eqref{eq:RB} is set to 2.34, not the value of the enriched atmosphere.
%Although it is a difficult problem which to use, our choice corresponds to the minimum value of $\RB$ and, thus, the atmospheric mass.}
In reality, the enriched (high-$\mu$) atmospheric gas diffuses out of the Bondi sphere by turbulence and part of it is replaced with unenriched disc gas that passes through the Hill sphere. 
High resolution hydrodynamic simulations will be needed to determine the residence time of the atmospheric gas leaking from the Bondi sphere in the Hill sphere and the resultant composition of gas around the Bondi radius, which is beyond the scope of this study. 
Instead, we choose the lower limit of $\RB$ by setting $\mu$ to 2.34. 
In this sense, the atmospheric mass we obtain in this study is a lower limit.

We assume that the gas density and temperature at the outer boundary of the atmosphere are equal to those of the surrounding disc gas and the disc gas properties are similar to those of the minimum-mass solar nebula \citep[MMSN;][]{Hayashi1981}; namely, the gas density is given by
\begin{equation}
  \rhoout = \num{1.5e-9} f_g
                \qty( \frac{a}{1~\si{AU}} )^{-9/4}
                \qty( \frac{M_*}{M_\odot} ) \, \si{g.cm^{-3}},
                \label{eq:rhoB}
\end{equation}
where $f_g$ is a disc gas depletion factor, and the temperature is
\begin{equation}
    \Tout = 280 \qty( \frac{a}{1~\si{AU}} )^{-1/2}
            \qty( \frac{L_*}{L_\odot} )^{1/4} \, \si{K}, \label{eq:TB}
\end{equation}
where $L_*$ is the stellar luminosity. In this study the values of $L_\ast$ are taken from the grid models for the pre-main-sequence (MS) and MS evolution of a star with constant mass 0.3~$M_\odot$ \citep{Baraffe+1998,Ramirez+Kaltenegger2014}. Note that recent models of pre-MS stellar evolution predict lower luminosities \citep[e.g.,][]{Kunitomo+2017}. 
Also, in protoplanetary discs, small dust particles are likely floating, preventing direct stellar irradiation \citep[e.g.,][]{Garaud+Lin2007}.
Thus, the above choice of $L_\ast$ and Eq.~\eqref{eq:TB} may lead to an overestimate of $\Tout$ and thus an underestimate of atmospheric mass.

% \remove{where we have assumed the stellar mass and luminosity relation as $L_* \propto M_*^2$, which is a typical mass-luminosity relation of low-mass pre-main sequence stars~\citep{Ramirez+Kaltenegger2014}.}
% \remove{
% which is obtained by fitting of observational data for main sequence stars of mass $\gtrsim 0.3M_\odot$ \citep{Hillenbrand+White2004,Scalo+2007}.
% }
% \kimura{
% Although the stellar luminosity significantly changes during the pre-main sequence phase~\citep{Baraffe+1998}, we use this simple formula because how the disk gas temperature is determined is complicated~\citep{Garaud+Lin2007, Oka+2011,Ida+2016} and also the stellar evolution in this phase is still uncertain~\citep{Kunitomo+2017}.
% }

% \kimura{
% We integrate Eqs.~\eqref{eq:hydrostat}--\eqref{eq:dTdr} from the smaller of $\RB$ and $\RH$ to the solid planet radius $\Rc = (3\Msolid/4\pi \rhos)^{1/3}$ for given $\Msolid$, $L$ and $\XHHO$.
% }

\section{Results}\label{Sec:Results}
First, we investigate the effects of water vapour enrichment on the atmospheric structure and mass with focus on a Mars-mass solid protoplanet.
We integrate Eqs.~\eqref{eq:hydrostat}--\eqref{eq:dTdr} 
inwards from the outer boundary to the inner one to obtain hydrostatic equilibrium solutions
%from the smaller of $\RB$ and $\RH$ to the solid planet radius $\Rc = (3\Msolid/4\pi \rhos)^{1/3}$ 
for given $\Msolid$, $L$ and $\XHHO$. 
%\remove{Boundary conditions are determined for given semi-major axis using Eqs.~\eqref{eq:rhoB} and \eqref{eq:TB}, assuming no orbital migrations.}
In Sect.~\ref{Sec:structure} we describe the basic properties of the enriched atmosphere and then show the sensitivity to disc gas density and homopause altitude in Sects.~\ref{sec:depletion} and \ref{sec:homopause}, respectively, and the effect of precipitation of the condensates in Sect.~\ref{sec:pseudo}.
Finally, in Sect.~\ref{sec:parameter study}, we investigate the captured water mass for various choices of the mass and semi-major axis of the protoplanet. 
Here we assume that the central star is %\remove{an M dwarf}
a pre-MS star of mass 0.3~$M_\odot$. %---
The results 
below do not explicitly depend on stellar mass,
since the outer boundary is determined by $\RB$ in all cases. 
In this section,
the stellar luminosity is assumed to be $0.3L_\odot$, which corresponds to the luminosity of a $0.3M_\odot$ pre-MS star at $\sim$1~Myr in the stellar evolution models of \citet{Baraffe+1998}.
%\citep{Baraffe+1998,Ramirez+Kaltenegger2014}}.

\subsection{Basic Properties of Enriched Atmosphere} %%%%%%%%%%%%%%%%%%%%%%%%%%%%%%%%%%%%
\label{Sec:structure}
\begin{figure}
    \centering
    \includegraphics[width=1.0\columnwidth]{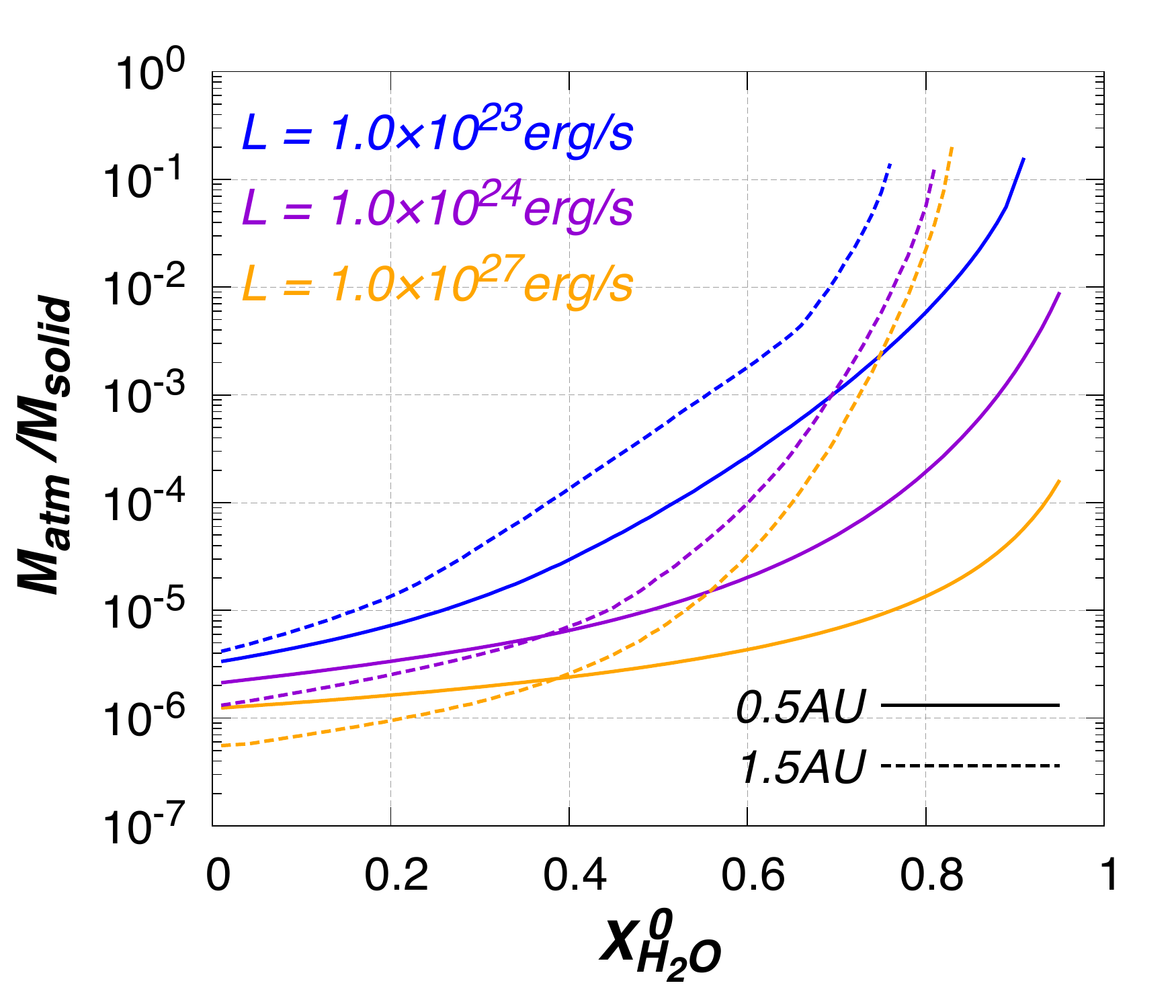}
    \caption{The atmospheric mass, $\Matm$, relative to the solid planet mass, $\Msolid$ (= 0.1~$M_\oplus$), as a function of the water mass fraction in the atmospheric gas, $\XHHO$, for 0.5~AU (solid lines) and 1.5~AU (dashed lines).
    The lines are colour-coded according to the assumed energy flux, $L$; the blue, purple, and orange ones are for $L= 1 \times 10^{23}$, $1 \times 10^{24}$ and $1 \times 10^{27}$~erg/s, respectively. }
    % \remove{Note that the purple and orange dashed lines are both drawn but overlapped almost completely.}}
    \label{fig:Menv-XH2O}
\end{figure}
\begin{figure}
    \centering
    \includegraphics[width=\columnwidth]{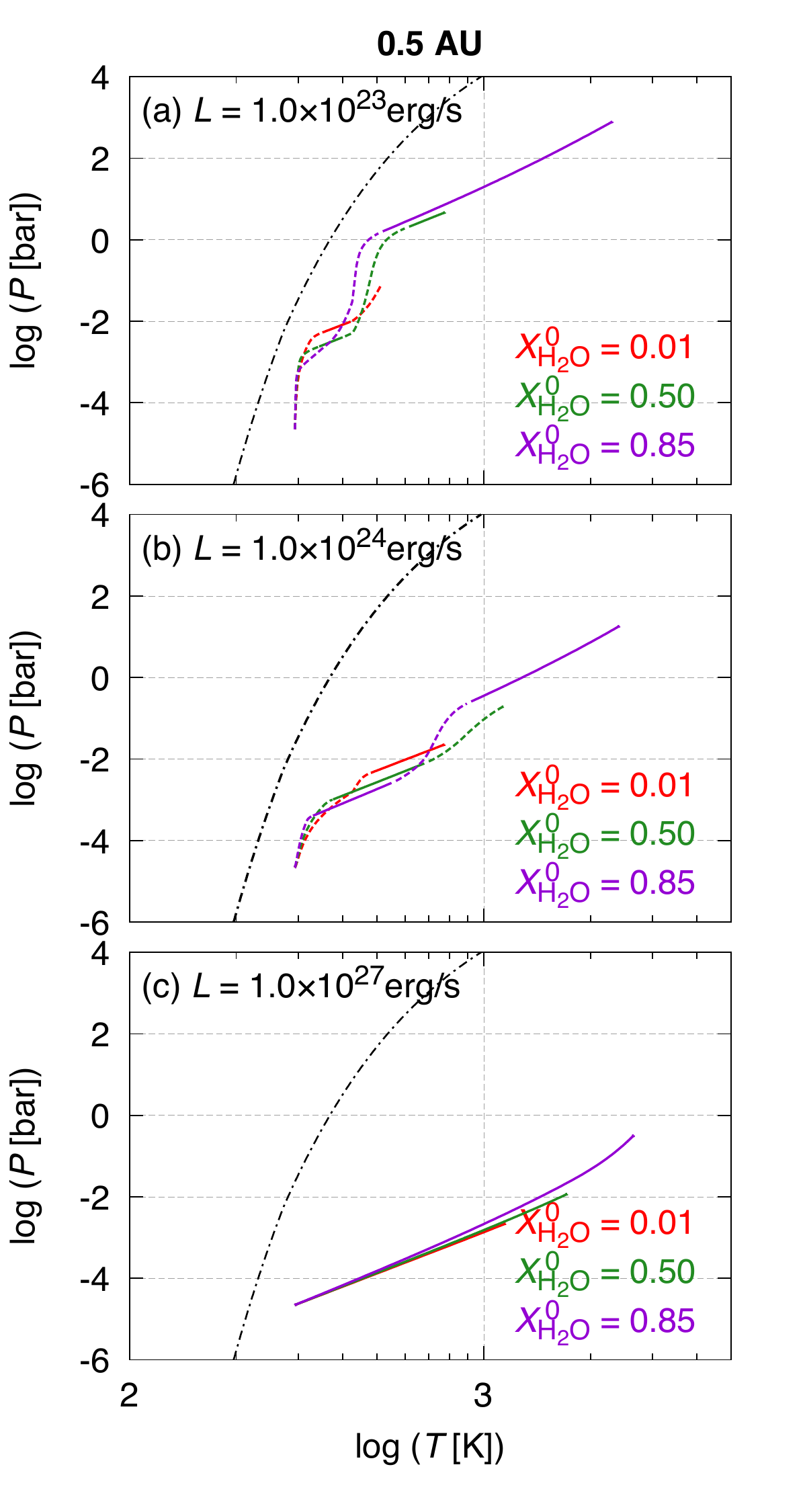}
    \caption{
    Temperature ($T$) vs. pressure ($P$) profile in the atmosphere for the semi-major axis of 0.5~AU with the energy flux $L$ of
    (a) $\SI{1.0e23}{erg/s}$, (b) $\SI{1.0e24}{erg/s}$, and (c) $\SI{1.0e27}{erg/s}$.
    The red, green, and purple lines are those for the water mass fraction $\XHHO=0.01$, 0.5, and 0.85, respectively.
    The solid and dashed lines indicate the convective and radiative regions, respectively.
    The black dot-dashed line indicates the saturation vapour pressure given by Eq.~\eqref{eq:psat}.
    }
    \label{fig:PT_model1}
\end{figure}
\begin{figure}
    \centering
    \includegraphics[width=\columnwidth]{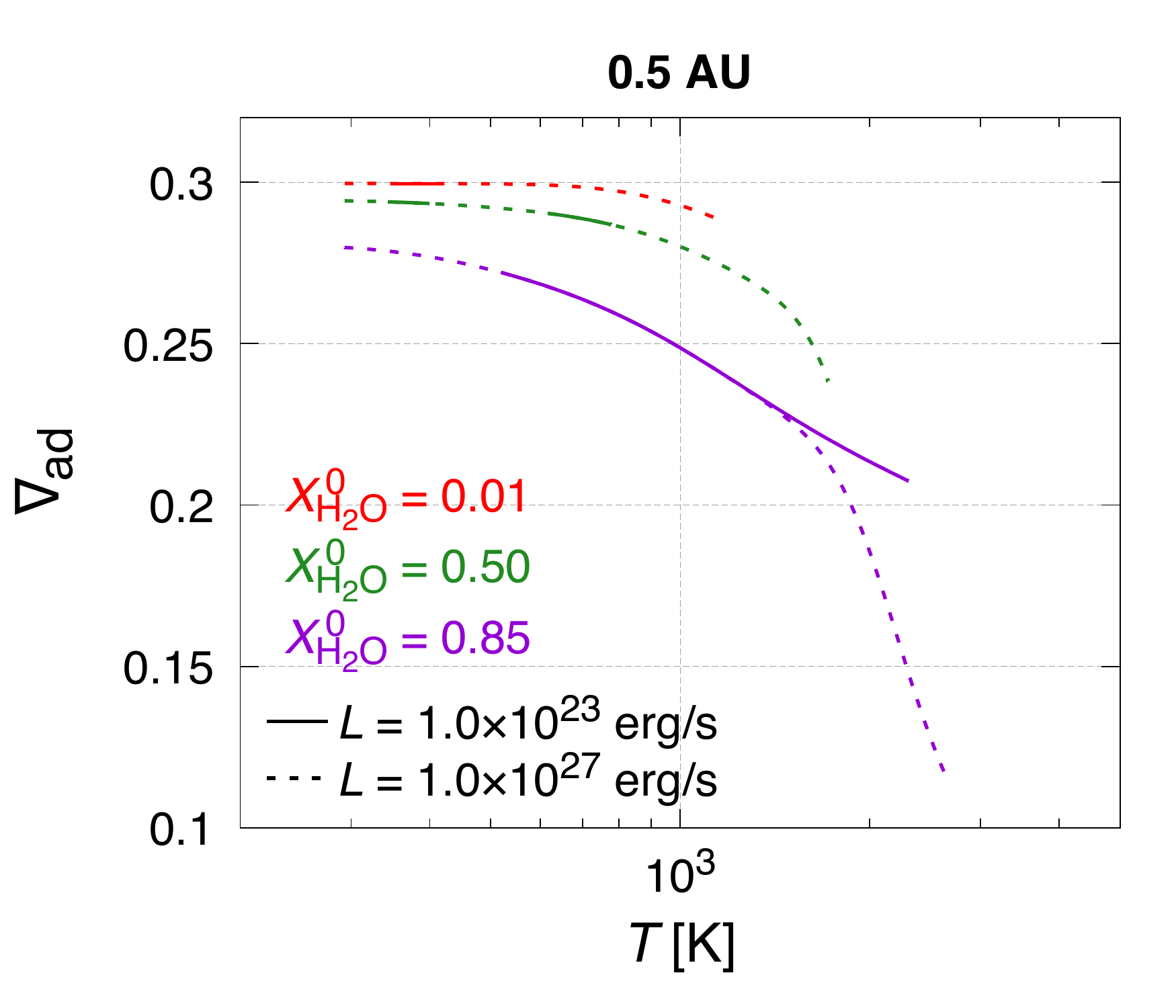}
    \caption{The adiabatic temperature gradient $\nabla_\mathrm{ad}$ in the convective zone as a function of temperature $T$ for the semi-major axis $a$ = 0.5~AU and
    the energy flux $L=\SI{1.0e23}{erg/s}$ (solid lines) and $L=\SI{1.0e27}{erg/s}$ (dashed lines).
    The red, green, and purple lines represent the results for the water mass fraction $\XHHO$ = 0.01, 0.5, and 0.85, respectively.}
    %\remove{Note that the red solid line is not plotted since the atmosphere is fully radiative.}}
    \label{fig:Ad_model1}
\end{figure}

\begin{figure}
    \centering
    \includegraphics[width=\columnwidth]{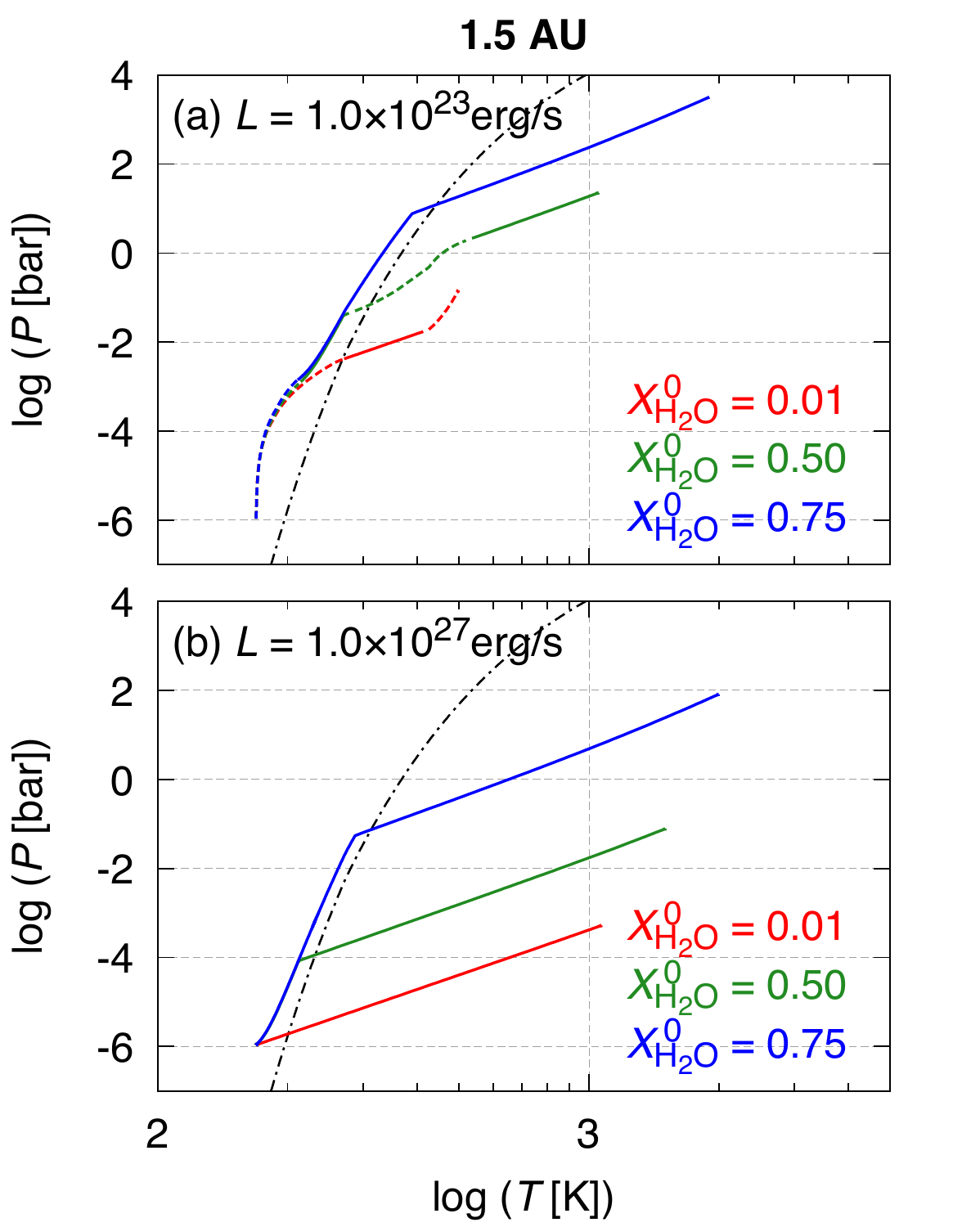}
    \caption{ 
    Same as Fig.~\ref{fig:PT_model1}, but for the semi-major axis of 1.5~AU with the energy flux $L$ of (a) $\SI{1.0e23}{erg/s}$ and (b) $\SI{1.0e27}{erg/s}$. 
    Note that the choice of $\XHHO$ is different from Fig.~\ref{fig:PT_model1}: the blue lines show the case with $\XHHO=0.75$.} 
    \label{fig:PT_model2}
\end{figure}

\begin{figure*}
    \centering
    \includegraphics[width=1.7\columnwidth]{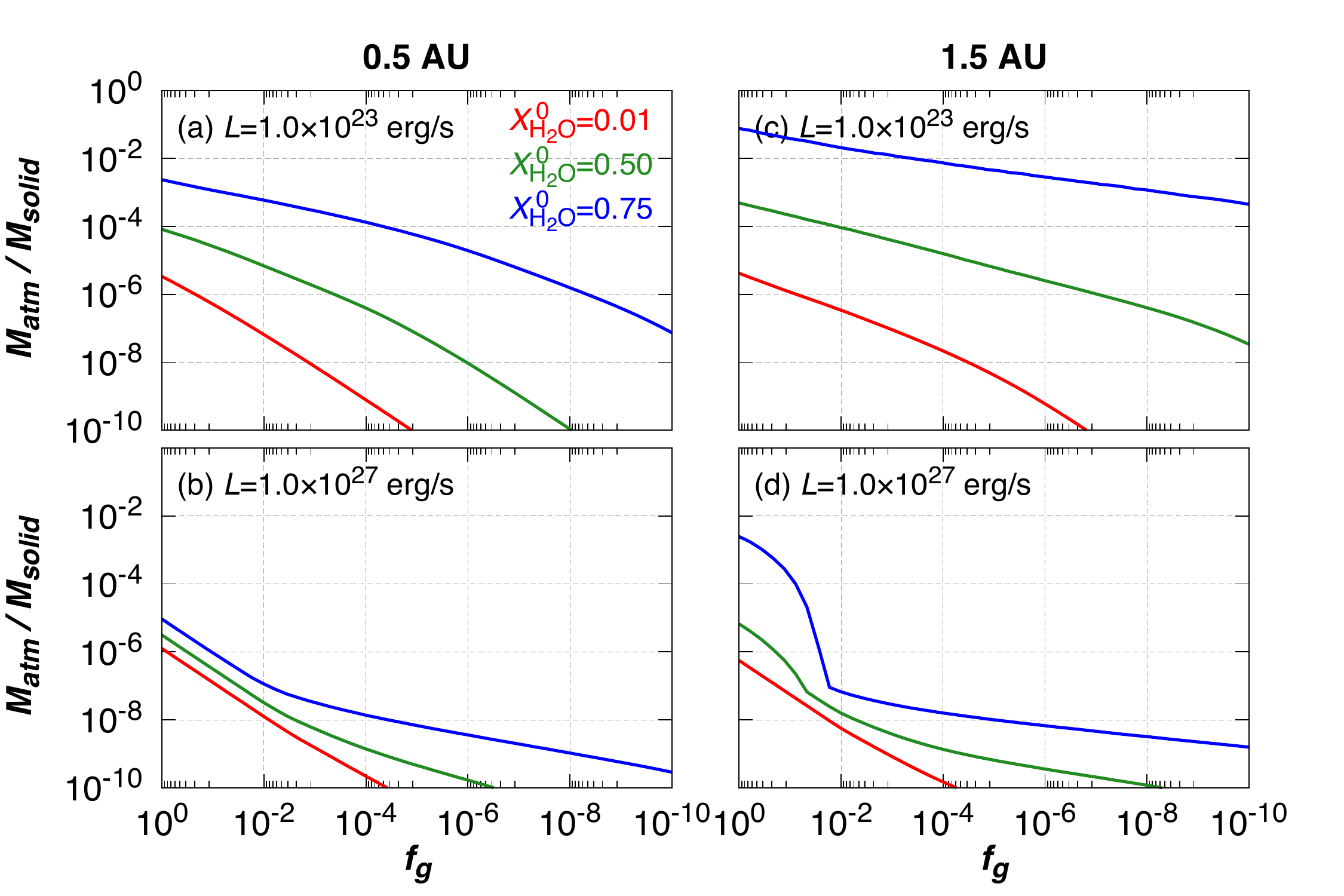}
    \caption{The atmospheric mass $\Matm$ (relative to the solid protoplanet mass $\Msolid$ = 0.1~$M_\oplus$) as a function of the disc gas depletion factor $f_g$ (see Eq.~\eqref{eq:rhoB} for the definition) for different choices of semi-major axis and energy flux as indicated in the panels.
    The red, green, and blue lines show the cases of $\XHHO=0.01, 0.5$ and $0.75$, respectively.}
    \label{fig:frho_Menv}
\end{figure*}

Figure~\ref{fig:Menv-XH2O} shows the relation between the atmospheric mass, $\Matm$ (relative to the solid protoplanet mass, $\Msolid$ = 0.1~$M_\oplus$) and the water mass fraction in the atmospheric gas, $\XHHO$, for three choices of the energy flux, $L$, and two choices of the semi-major axis, $a$.
Here we have chosen 0.5 and 1.5~AU for the semi-major axes to examine both cases where water vapour condenses in the atmosphere and where it does not.
% \remove{
% since the location of the conventional circum-stellar habitable zone around main-sequence stars of 0.3~$M_\odot$ is estimated at $0.12$--$0.23$~AU\citep{Kopparapu+2014}.
% }
The three choices of $L=\num{1e23},\num{1e24}$, and $\SI{1e27}{erg/s}$ correspond to the planetesimal accretion rates $\dot{M}_\mathrm{solid}$ of 
$\sim 5 \times 10^{-9}$, $5 \times 10^{-8}$, and $5 \times 10^{-5} M_\oplus\si{/yr}$, respectively,
%\kimura{
which are obtained by the relation
\begin{equation}
    L = \frac{G\Msolid \dot{M}_\text{solid}}{\Rc}.
\end{equation}
%\remove{where $\dot{M}_\text{solid}$ is the planetesimal accretion rate.} 
While the solid accretion process remains a matter of debate, standard theories of planetesimal accretion \citep[e.g.,][]{Kokubo+Ida2002} estimate that the
typical %\remove{planetary accretion rate} 
value of $\dot{M}_\mathrm{solid}$ is on the order of $10^{-5}$~$M_\oplus$/yr for protoplanets of 0.1~$M_\oplus$ at 0.5~AU in a protoplanetary disc with a MMSN-like density profile around a 0.3~$M_\odot$ star.
%\remove{for planet with $0.1\Msolid$ at 0.1AU around $0.3M_\odot$ is estimated about $\sim 10^{-5}M_\oplus/\si{yr}$ in the case with the disk gas and solid surface density of MMSN model~\citep{Kokubo+Ida2002}.}
%\remove{However, it} 
The accretion rate declines as the disc gas dissipates or the solid surface density decreases.
%\remove{Therefore,} 
Thus, we also consider low values of
%\remove{use a wide range of} 
$L$ here. 
%\remove{as an input parameter}.
%}

An overall trend is that $\Matm$ increases with $\XHHO$. 
Especially for high $\XHHO$ ($\gtrsim$~0.5), water enrichment has such a significant impact on the atmospheric structure that $\Matm$ increases by a few orders of magnitude. 
Also, $\Matm$ is larger for lower $L$, as previously known \citep{Ikoma+Genda2006}.
The sensitivity of $\Matm$ to $\XHHO$ differs depending on $L$ and $a$.
For small values of $\XHHO$,
the lower the energy flux, the steeper the slope of the curve is. 
Consequently, for large values of $\XHHO$, the atmospheric mass for $L$ = $\SI{1.0e23}{erg/s}$ is much larger than those for $L = 1 \times 10^{24}$ and $10^{27}$~erg/s,
except that $\Matm$ is weakly dependent on $L$ for $\XHHO \gtrsim 0.7$ at $a$ = 1.5~AU (dashed lines).
The dependence of $\Matm$ on $\XHHO$ itself is greater at $a$= $1.5~\si{AU}$ than at $a = 0.5~\si{AU}$.

The above features can be interpreted as follows. 
Figure~\ref{fig:PT_model1} shows the temperature vs. pressure profiles in the atmosphere with three different values of $\XHHO$ calculated at $a$ = 0.5~AU for $L$ = $\SI{1.0e23}{erg/s}$ (a), $\SI{1.0e24}{erg/s}$ (b), and $\SI{1e27}{erg/s}$ (c).
In panel~(a), because of low energy flux (and low gravity), the atmosphere is almost entirely radiative for $\XHHO$ = 0.01 (and actually $\XHHO < 0.5$, although not shown). 
Thus, the main reason for the increase in atmospheric mass is an increase in mean molecular weight, which dominates over the effect of enhanced H$_2$O opacity, although narrow convective zones exist in some cases. 
Note that the calculated grain depletion factor $f$ is as small as $\sim$~0.001.
In the case of $L = 1 \times 10^{23}$~erg/s and $a = 0.5$~AU, 
(also see the blue solid line of Fig.~\ref{fig:Menv-XH2O}),
the mean molecular weight $\mu$ increases from 2.3 to 3.6  as $\XHHO$ increases from 0.01 to 0.4.
Even such a small increase in $\mu$ yields a large increase in $\Matm$ \citep[see also][for an analytical interpretation]{Stevenson1982}.
In contrast, for $\XHHO \gtrsim 0.5$, a convective zone appears in the deep atmosphere, which accounts for a large fraction of the atmospheric mass, so that a decrease in the adiabat $\NBLad$ causes the increase in atmospheric mass, in addition to mean molecular weight (see solid lines in Fig.~\ref{fig:Ad_model1} where $\NBLad$ is shown as a function of temperature in the convective zone).
The decrease in $\NBLad$ is due to the increase in the specific heat caused by dissociation of some molecules such as H$_2$O \citep[see also][]{Hori+Ikoma2011}.
Therefore the atmospheric mass increase is much steeper for high $\XHHO$ than for low $\XHHO$.

As shown in Fig.~\ref{fig:PT_model1}(c), for $L=\SI{1.0e27}{erg/s}$ (and actually $L \gtrsim \SI{1.0e25}{erg/s}$), dry convection occurs entirely in the atmosphere, regardless of $\XHHO$.
Similarly to radiative atmospheres, the atmospheric mass depends strongly on mean molecular weight \citep[see also][for an analytical interpretation]{Wuchterl1993,Ikoma+2001}.
The steeper increase in atmospheric mass for $\XHHO \gtrsim 0.8$ is due to a decrease in $\NBLad$, similarly to the case with $L=\SI{1.0e23}{erg/s}$ (see dashed lines in Fig.~\ref{fig:Ad_model1}).

For $L$ = $\SI{1.0e24}{erg/s}$ and $a$ = 0.5~AU, the dependence of $\Matm$ on $\XHHO$ is similar to but greater than that of $L=\SI{1.0e27}{erg/s}$ for $\XHHO \gtrsim 0.6$.
This is because the atmosphere is partly radiative for $L = \SI{1.0e24}{erg/s}$ (see Fig.~\ref{fig:PT_model1}(b)); 
the pressure increase in this radiative region makes the inner convective region denser, resulting in a more massive atmosphere.

Finally, as described above, $\Matm$ increases more rapidly with $\XHHO$ for $a$ = 1.5~AU than for $a$ = 0.5~AU.
Figure~\ref{fig:PT_model2} shows the $P$-$T$ profiles for $a=$1.5~AU.
%\kimura{
Note that the results with $\XHHO=0.75$ are shown as the most enriched case, in contrast to Fig.~\ref{fig:PT_model1} and \ref{fig:Ad_model1}, because there is no hydrostatic solution for $\XHHO>0.75$ in the case with $a=1.5$~AU and $L = \SI{1e23}{erg/s}$.
%}
Regardless of energy flux, moist-convection dominates energy transfer in the upper part of the atmosphere with a high $\XHHO$. 
In those moist-convective regions, pressure increases significantly while temperature increases a little, in contrast to dry convective regions, as found from the comparison between Figs.~\ref{fig:PT_model1} and \ref{fig:PT_model2}. 
In Fig.~\ref{fig:PT_model2}, the convection is switched from the moist to dry one when the partial pressure of vapour becomes lower than the saturation vapour pressure. 
Since the dry-adiabat hardly changes with water fraction, the surface pressure (or the pressure at the bottom of the atmosphere) and the atmospheric mass are almost determined by the pressure at the switching point.
A slight increase in $\XHHO$ is found to result in a large increase in the switching-point pressure, which leads to a sudden increase in the atmospheric mass.

\subsection{Effect of Disc Gas Depletion}
\label{sec:depletion}

\begin{figure}
    \centering
    \includegraphics[width=\columnwidth]{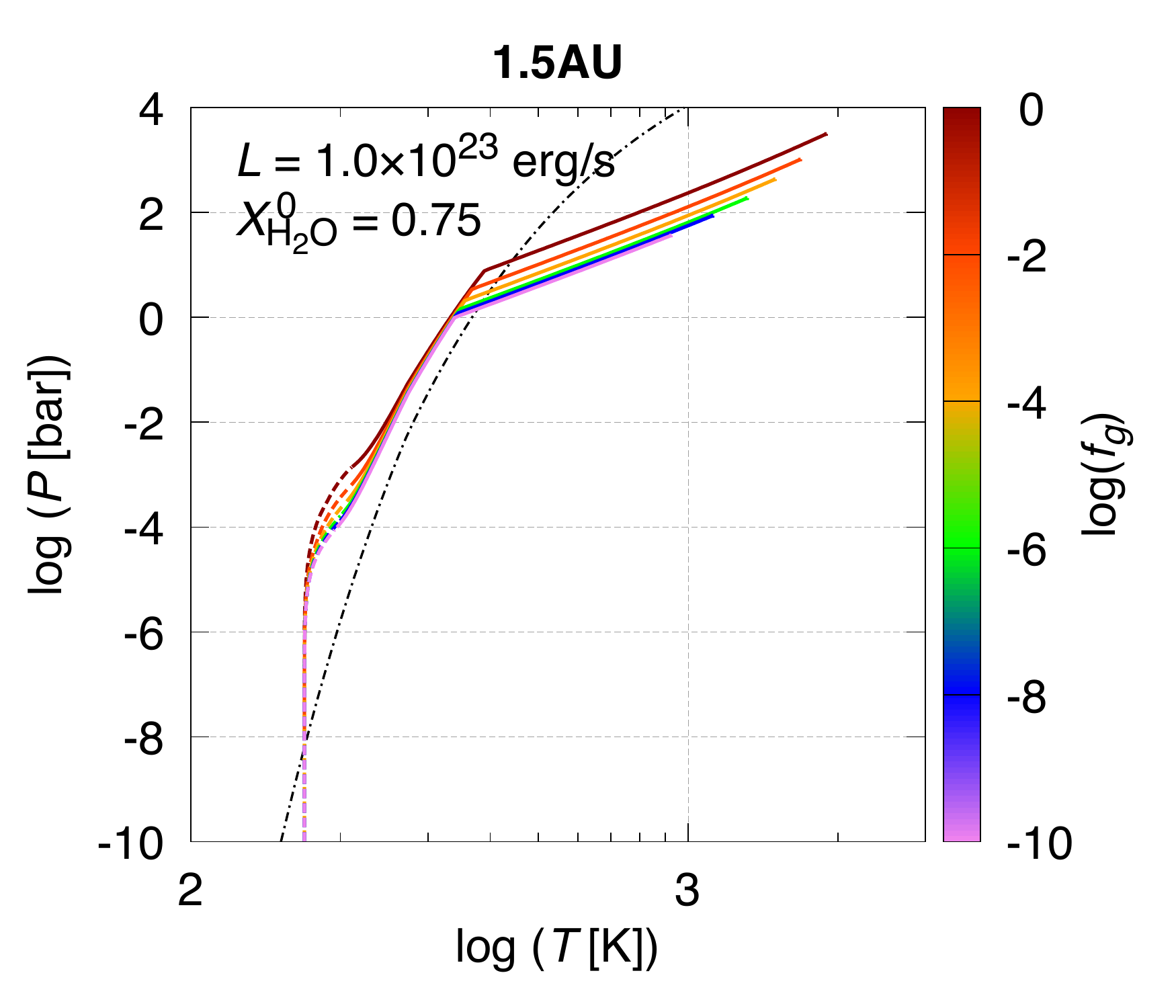}
    \caption{Temperature ($T$) vs. pressure ($P$) profiles in the enriched atmosphere for different values of the disc depletion factor $f_g$ (see Eq.~(\ref{eq:rhoB}) for the definition of $f_g$). 
    Here we have assumed that the water fraction $\XHHO$ is 0.75, the semi-major axis is 1.5~AU and the energy flux is $\SI{1e23}{erg/s}$.
    The lines are colour-coded according to $\log f_g$ (= $-10$, $-8$, $-6$, $-4$, $-2$, and 0).
    The solid and dashed lines represent the convective and radiative region, respectively.
    The dot-dashed line indicates the saturation vapour pressure.}
    \label{fig:Model2_lowL_highX}
\end{figure}
\begin{figure}
    \centering
    \includegraphics[width=\columnwidth]{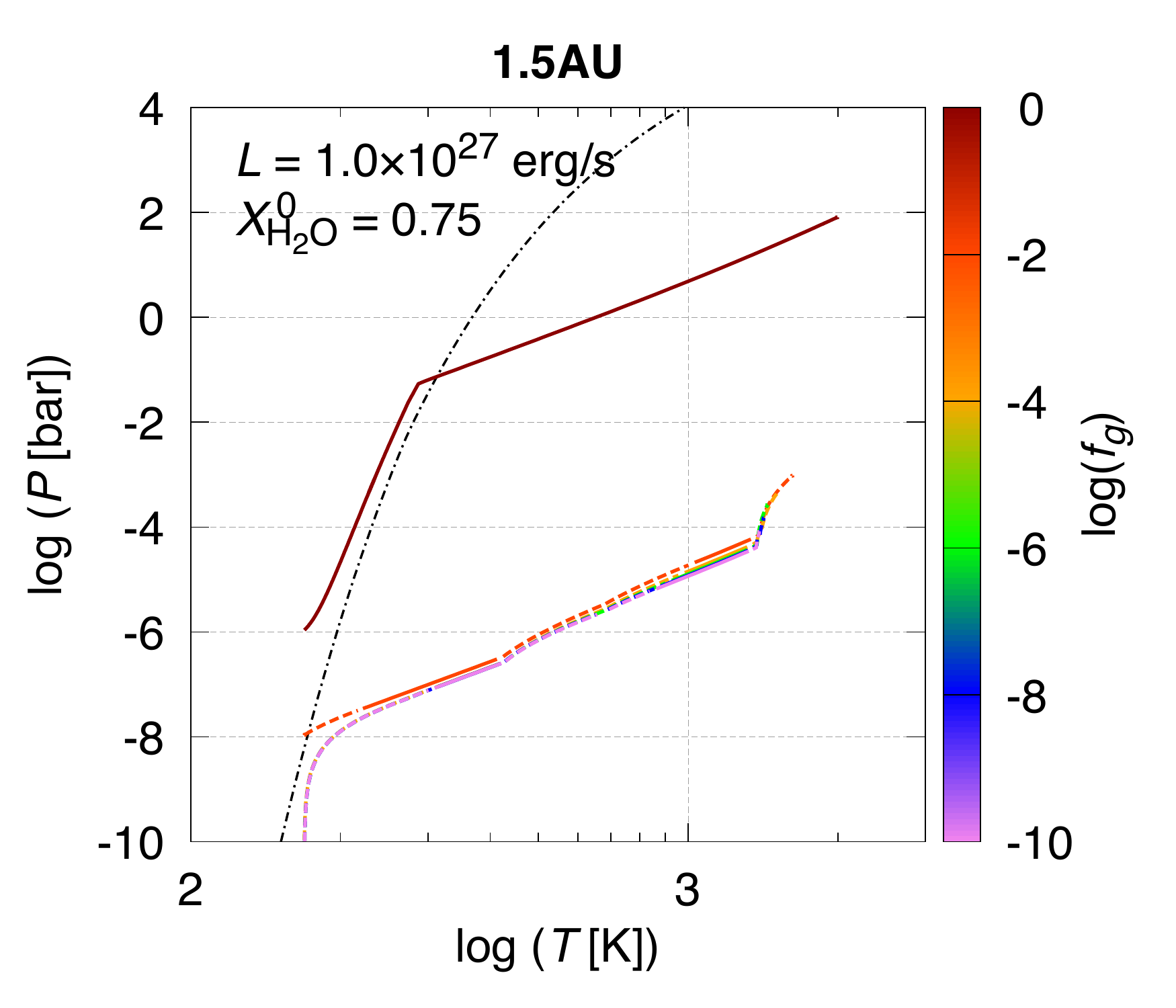}
    \caption{Same as Fig.~\ref{fig:Model2_lowL_highX}, 
            but for the energy flux $L=\SI{1e27}{erg/s}$}.
    \label{fig:Model2_highL_highX}
\end{figure}

The atmosphere of this type reduces its mass, as the disc gas density declines; however, the amount of reduction in atmospheric mass depends on the strength of gravitational binding \citep{Ikoma+Genda2006,Stokl+2015}. 
Here we investigate quantitatively how water enrichment affects this trend by varying the disc gas depletion factor $f_g$ (see Eq.~\eqref{eq:rhoB}) from $1$ to $1 \times 10^{-10}$.
The solid planet mass is fixed to $0.1M_\oplus$, the same as the previous section.
The results are shown in Fig.~\ref{fig:frho_Menv}.

An obvious trend is that the mass of the enriched atmosphere decreases more slowly with decreasing $f_g$ than that of the non-enriched atmosphere ($\XHHO$ = 0.01), except for $f_g \lesssim 10^{-2}$ for $\XHH$ = 0.75 and $L$ = $\SI{1e27}{erg/s}$.
Especially in the case of high enrichment, low energy flux, and low disc gas temperature ($\XHHO$ = $0.75$, $L$ = $\SI{1e23}{erg/s}$, and $a$ = 1.5~AU), the atmospheric mass is still larger than $\sim$ $10^{-4}\Msolid$ ($\simeq 10^{-5} M_\oplus$) even after the severe disc gas depletion ($f_g \sim 10^{-10}$); this water mass is comparable to the total mass of the Earth oceans. 
This result is obviously different from the conclusion of \citet{Ikoma+Genda2006} that the non-enriched primordial atmosphere of a Mars-mass protoplanet is quite sensitive to the outer boundary density and, thus, never survives disc gas dispersal. 
This is not always the case, however;
as shown in Fig.~\ref{fig:frho_Menv}(b) and (d), for $L=\SI{1.0e27}{erg/s}$, atmospheric mass decreases by more than $\sim 5$ orders of magnitude even for $\XHHO=0.75$ until $f_g = 1 \times 10^{-10}$.
% \remove{Furthermore, the atmospheric mass dependence on $f_g$ is different between $a=$0.1AU and 0.2AU, especially for high-$\XHHO$.}
% \remove{For $a$ = 0.2~AU, $\Matm/\Mc$ is $>10^{-2}$ with $\XHHO=0.75$ for $f_g=1$, but it decreases drastically around $f_g\sim 10^{-2}$, and the planet cannot keep sufficient amount of atmosphere after the disk gas depletion.}

Again, such different features can be interpreted from the atmospheric structure, as follows.
First, as found in \citet{Ikoma+Genda2006}, the non-enriched atmosphere on a Mars-mass protoplanet is wholly radiative and so thin as to be vulnerable to a change in boundary conditions.
Consequently, the atmospheric mass decreases almost linearly with decreasing $f_g$ (see Fig.~\ref{fig:frho_Menv}).

In contrast, the decrease in the mass of the enriched atmosphere is less significant, regardless of the protoplanet's location.
Figure~\ref{fig:Model2_lowL_highX} shows $T$-$P$ profiles for different values of $f_g$ in the case of $\XHHO$ =0.75, $a$ = 1.5~AU and $L=\SI{1e23}{erg/s}$.
Even for low $f_g$, gas density increases so rapidly in the outer isothermal layer that water condensation occurs.
As described in Sec.~\ref{Sec:structure}, the mass of the atmosphere with an outer moist-convective region is determined mostly by the pressure at the switching point from the moist to dry convection.
Since this switching-point pressure is insensitive to the boundary conditions, the dependence of highly enriched atmospheric mass on $f_g$ becomes quite weak relative to the non-enriched atmosphere.
% Even in the warm circumstellar region ($a=0.1$AU), where no condensation occurs, the outer radiative region in the highly enriched case is so thick 
% that the effect of the boundary conditions on the atmospheric mass is small.

In the high-luminosity case, however, the atmospheric mass depends greatly on $f_g$ even for high $\XHHO$ (see Fig.~\ref{fig:frho_Menv}(b) and (d)).
The sudden change in the slope of the blue line in Fig.~\ref{fig:frho_Menv}(b) and (d) around $f_g\sim 10^{-2}$ is caused by the shift of the dominant atmospheric structure from convective to radiative.
The drastic decrease at $f_g\sim 10^{-2}$ of the blue line in Fig.~\ref{fig:frho_Menv}(d)  appears because of the end of water vapour saturation (see Figure~\ref{fig:Model2_highL_highX})

\subsection{Effect of homopause location} \label{sec:homopause}
\begin{figure}
    \centering
    \includegraphics[width=\columnwidth]{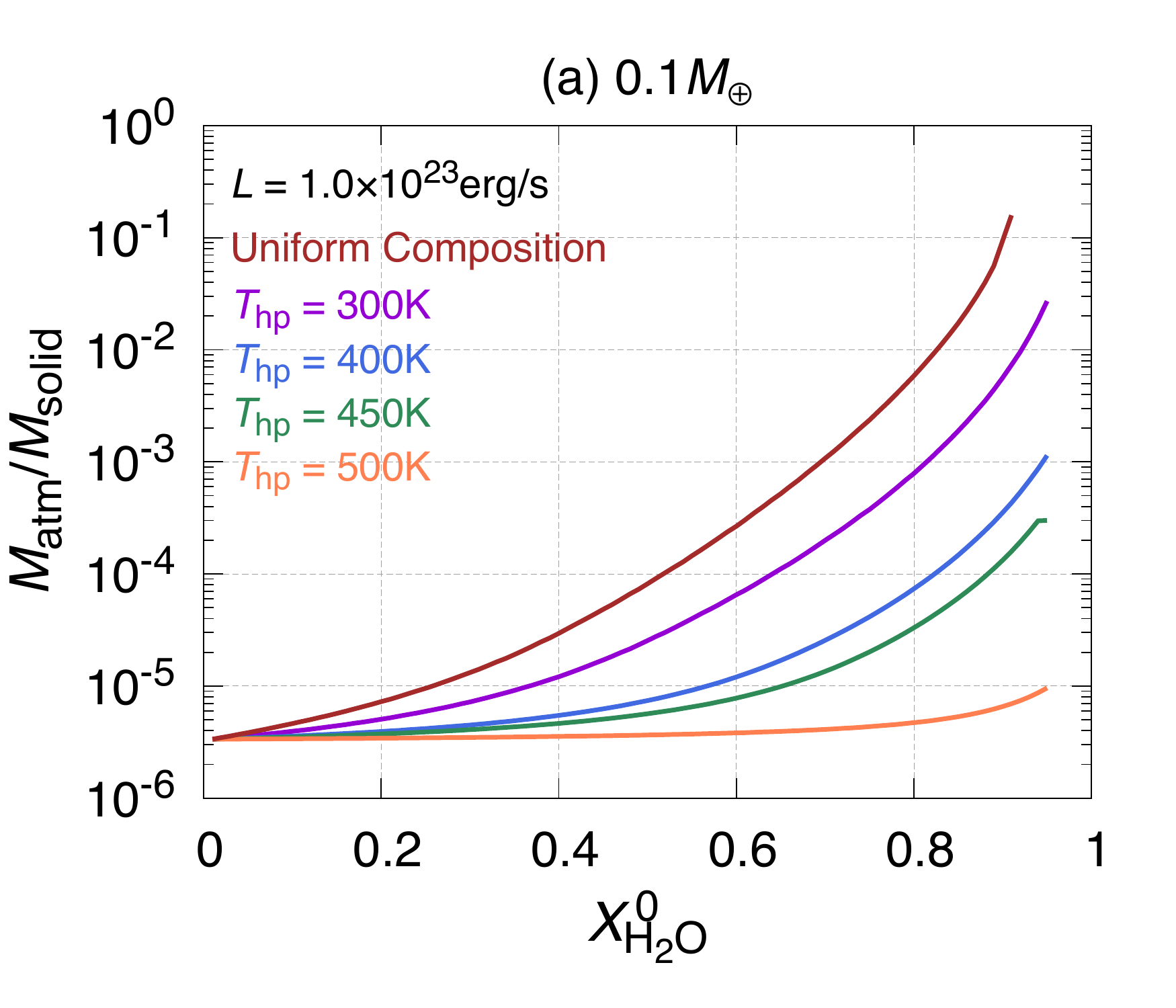}   
    \includegraphics[width=\columnwidth]{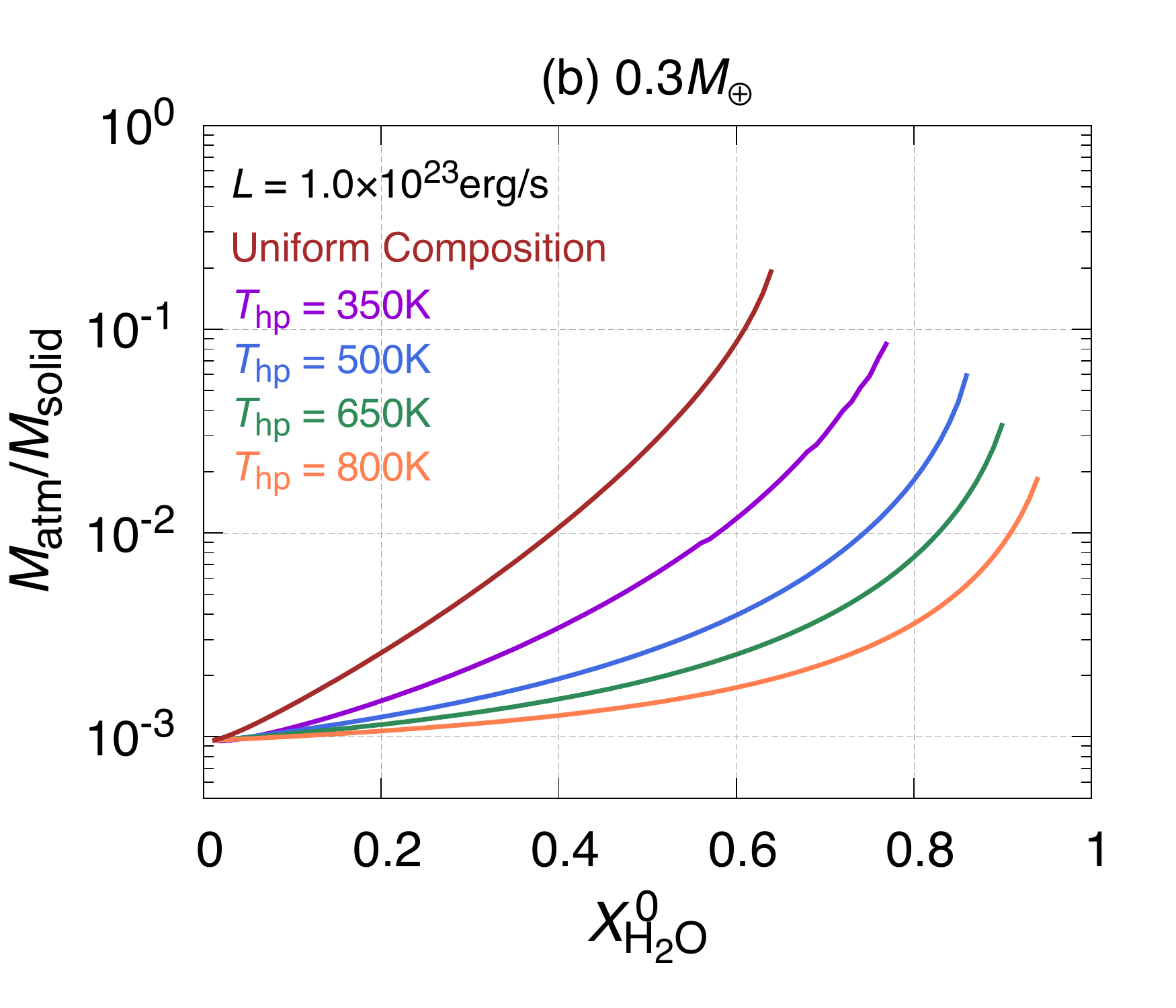}
    \caption{Dependence of the atmospheric mass on the homopause temperature. The atmospheric mass relative to the solid protoplanet mass of 0.1~$M_\oplus$ (a) and 0.3~$M_\oplus$ (b) is shown as a function of the water mass fraction in the lower, enriched layer of the atmosphere, $\XHHO$ in the case of 0.5~AU and $L=\SI{1e23}{erg/s}$.
    The lines are colour-coded according to the homopause temperature $T_\mathrm{hp}$, as indicated in the panels.
    The brown line shows the results for the uniform composition case.}
    \label{fig:xH2O_Menv_Thp}
\end{figure}

We have assumed so far that the composition is uniform throughout the atmosphere. 
In reality, however, the atmosphere may not always be uniform (see Sect.~\ref{sec:discussion_uniformity}).
Here we consider a compositionally two-layered atmosphere, namely, a non-enriched ($\XHHO$ = 0.01) upper layer on top of an enriched lower one.
Hereafter the boundary between the two layers is called the homopause.
%\kimura{
We parameterize the homopause location with its temperature $T_\text{hp}$.
We perform inward integration %is performed 
assuming $\XHHO=0.01$ for $T < T_\text{hp}$ and switch $\XHHO$
%is switched 
to the enriched value for $T > T_\text{hp}$.
%Similar approach is done in \cite{Hori+Ikoma2011}.
%}

Figure~\ref{fig:xH2O_Menv_Thp} shows the atmospheric mass as a function of the water mass fraction in the lower enriched layer of the atmosphere $\XHHO$ for different values of homopause temperature $T_\text{hp}$ and two choices of 
the solid protoplanet mass, 0.1~$M_\oplus$ (a) and 0.3~$M_\oplus$ (b); $a$ = 0.5AU and $L$ = $\SI{1e23}{erg/s}$.
It turns out that the presence of the upper non-enriched layer has a great impact on the atmospheric mass especially for high $\XHHO$. 
For both cases, atmospheric mass decreases, as $T_\text{hp}$ increases (i.e. as the non-enriched layer becomes deeper).
In the case of $0.1M_\oplus$, the atmospheric mass at $\XHHO=0.8$ for $T_\text{hp}=400$~K is lower by two orders of magnitude than that for the uniform composition.
Furthermore, no significant increase in atmospheric mass occurs for $T_\text{hp}$ = 500~K, because the surface temperature of the fully non-enriched atmosphere is $\sim$500~K for this parameter set.
In the case of $0.3M_\oplus$, in contrast, atmospheric mass increases with $\XHHO$ even for $T_\text{hp}=800$~K, because the surface temperature of the fully non-enriched atmosphere is much higher than that ($\sim$1000~K) in this case.
However, the atmospheric mass with $T_\text{hp}>500$~K decreases by more than one order of magnitude compared to the case of uniform composition.
Therefore, in any case, the homopause altitude has a large impact on the atmospheric mass, and thus, the water amount.

\subsection{Effect of precipitation of condensates}\label{sec:pseudo}%%%%%

We have assumed so far that the condensed water remains in the atmosphere and its specific heat contributes to the atmospheric heat budget.
However, that is not always the case; the droplets may precipitate quickly and are removed from the condensation region.
Here we investigate this effect on the atmospheric mass. 
To do so we set $\rl = 0$ in Eq.~\eqref{eq:moist-adiabat}.
Although, in reality, the water droplets would fall and then vaporise, enriching somewhat the inner, unsaturated regions \citep{Chambers2017}, we neglect such an effect, because the inner convective regions contain much more mass than the condensation regions.
Below we assume that the atmosphere is uniform in composition again.

\begin{figure}
    \centering
    \includegraphics[width=\columnwidth]{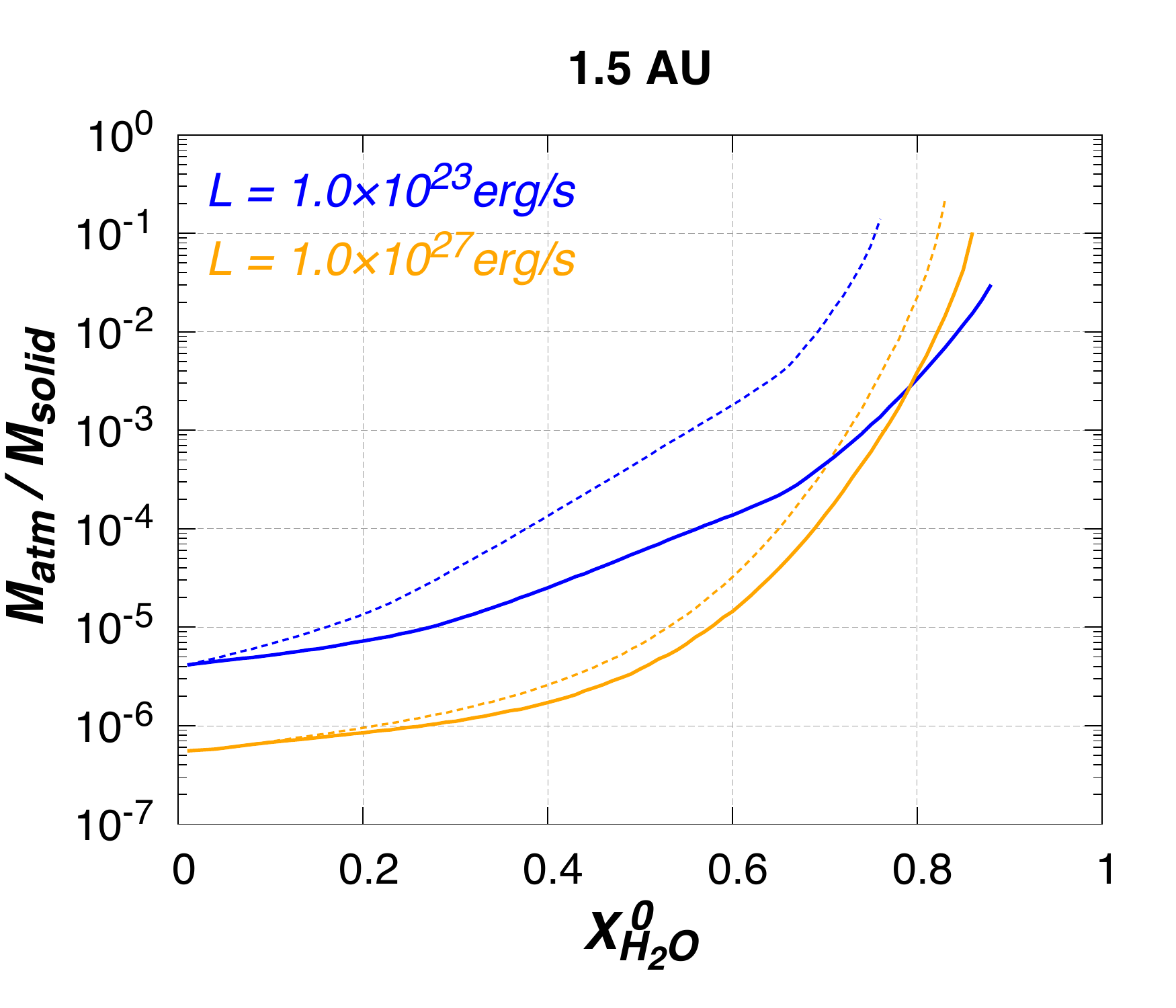}
    \caption{The atmospheric mass fraction as a function of $\XHHO$.
    The solid and dashed lines show the case with and without the precipitation, respectively.}
    \label{fig:xH2O_Menv_pseudo}
\end{figure}

Figure~\ref{fig:xH2O_Menv_pseudo} shows the calculated atmospheric mass as a function of $\XHHO$ for $a$ = 1.5~AU and $L$ = $\num{1e23}$ and $\SI{1e27}{erg/s}$ and compares the results with (solid lines) and without (dashed lines) precipitation. It is found that the effect of precipitation lowers the atmospheric mass.
For $\XHHO \lesssim 0.5$, the atmospheric mass differs by only about a factor of two.
However, for $\XHHO \gtrsim 0.7$, the difference becomes about an order of magnitude. 
We discuss whether precipitation likely occurs in Sect.~\ref{sec:discussion_precipitation}.

\subsection{Water amount of planets with various masses and semi-major axis}
\label{sec:parameter study}
%%%%%%%%%%%
\begin{figure*}
    \centering
    \includegraphics[width=2\columnwidth]{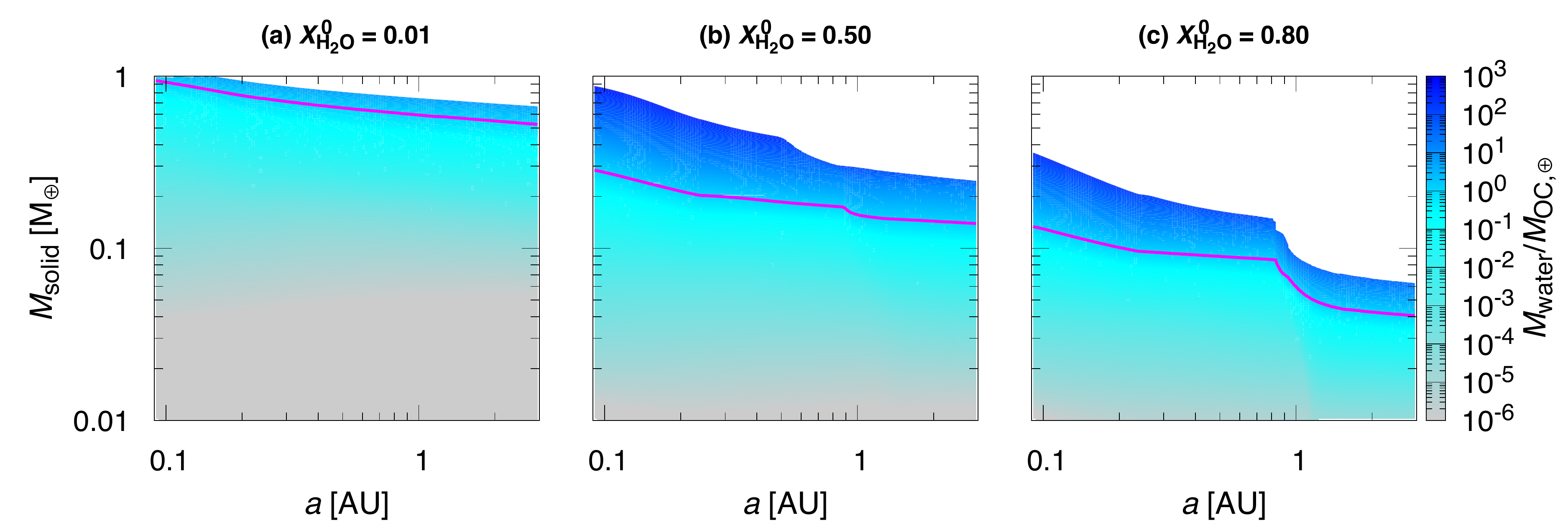}
    \caption{
     Colour contour plots of the captured water mass, $\XHHO \Matm$, as a function of the semi-major axis $a$ and the solid planet mass, $\Msolid$. The magenta line represents the set of $\Msolid$ and $a$ for which $\XHHO \Matm$ = 1~$M_\mathrm{OC,\oplus}$. 
     Here the energy flux $L=\SI{1.0e23}{erg/s}$, the disc depletion factor $f_g = 1.0$, and the stellar mass is $0.3M_\odot$.
     The captured water mass is shown in the unit of the Earth ocean mass ($M_{\rm OC, \oplus}$ = $\num{2.3e-4}M_{\oplus}$).
     There is no hydrostatic solution in the blank area.
    }
    \label{fig:Mw_before_loss}
\end{figure*}
We calculate the total mass of the captured water (i.e., $\XHHO \Matm$) for the protoplanet with various masses and semi-major axes.
Here we assume that the atmosphere is uniform in composition and the condensed water remains in the moist convective regions.
Figure~\ref{fig:Mw_before_loss} is colour-contour plots of $\XHHO \Matm$ for different three choices of $\XHHO$, 0.01 (a), 0.50 (b), and 0.80 (c) with $L=\SI{1.0e23}{erg/s}$ and $f_g=1.0$.
The magenta lines indicate the sets of $a$ and $\Msolid$ for which the captured water mass is equal to the Earth ocean mass. 
%\kimura{
Note that no %hydrostatic 
solution is found %for planets 
in the blank areas: %because they exceed the critical core mass. 
This is because the assumed value of $L$ is too small for the atmosphere to be in hydrostatic equilibrium. In reality, the atmosphere contracts quasi-statically and transforms its gravitational energy to thermal energy, leading to runaway accumulation of disc gas
%the atmospheres of such planets begin quasi-static contraction and accumulate additional disk gas~
\citep[e.g.,][]{Ikoma+2000}.
%}

Whereas the protoplanet has to be several times as massive as Mars to acquire water comparable in amount with the Earth's oceans in the case of $\XHHO=0.01$, 
less massive protoplanets with Mars mass and even sub-Mars mass suffice to do so in the case of $\XHHO=0.5$ and 0.8, especially in the relatively cool circumstellar regions.
The sudden changes in the water mass (and the threshold mass of the solid planet beyond which no hydrostatic solution exists) around 0.5--0.8~AU seen in Fig.~\ref{fig:Mw_before_loss}(b) and (c) are due to water condensation.
Then, farther than 1.0~AU, the contour lines are almost horizontal.
This is because the atmospheric structure is dominated by moist adiabat and thus insensitive to the boundary conditions (see Sect.~\ref{sec:depletion}). 
%\kimura{
Also, for planets at $\lesssim 0.8$~AU, the obtained water amount is quite insensitive to the semi-major axis and, thus, the disk temperature.
% Therefore, even though the disk gas temperature in the planet forming phase would be different from Eq.~\eqref{eq:TB}~\citep[e.g.,][]{Garaud+Lin2007,Oka+2011,Ida+2016}, we expect this result is hardly affected 
%}
\section{Discussion}%%%%%%%%%%%%%%%%%%%%%%%%%%%%%%%%%%%%%%%%
\label{Sec:Discussioin}
To know how much water terrestrial protoplanets of 0.01--1~$M_\oplus$ can acquire in situ, we have made detailed investigations on the structure and mass of the captured atmosphere that is enriched with water. 
Here we discuss our key assumptions and some ignored processes. 

\subsection{Enrichment with water}
What we have demonstrated in Sect.~\ref{Sec:Results} is that even a protoplanet of Mars-mass or smaller obtains water comparable in mass to the present Earth's oceans, as long as the atmosphere is highly enriched with water ($\gtrsim$~50~\% by weight or $\gtrsim$~10~\% by 
mole number). 
The question is whether such enrichment occurs on low-mass protoplanets. 
As mentioned in Introduction, once enriched with water (or other volatiles) and thereby becoming hot enough for rocks to be molten in some way, the atmosphere would keep itself enriched through chemical reactions between the atmospheric hydrogen and oxidising rocky materials \citep[][]{Sasaki1990,Ikoma+Genda2006}.

First, volatile-rich planetesimals would readily enrich the atmosphere with water: 
Even in shallow parts of the atmosphere, the temperature is high enough for ice to evaporate. 
An icy planetesimal of radius 100~km and density 2~g/cc, for example, has a mass of $\sim 1 \times 10^{-6} M_\oplus$.
Thus, 
%\kimura{
%although it would be difficult for planets around M dwarfs to obtain volatile-rich planetesimals as mentioned in Sec.~\ref{sec:intro},
%}
just a few such planetesimals are comparable in mass to the atmosphere with the solar abundances ($\XHHO$ = 0.01) on a Mars-mass solid protoplanet (see Fig.~\ref{fig:Menv-XH2O}). 
Although planets in the habitable zone around M dwarfs are unlikely to collide with many such planetesimals, planetary embryos and protoplanets in outer regions can scatter some icy planetesimals beyond the snowline inward occasionally by, for example, repeating scattering \citep{Raymond+2007}.
Not icy but chondritic rocky planetesimals could also enrich the atmosphere via impact degassing. 
Such degassing is known to occur on a protoplanet larger than the moon ($\gtrsim$ 0.01~$M_\oplus$) \citep[][]{Abe+86,Zahnle+88}.

Dry rocky planetesimals never emit water nor other blanketing-effect gases directly on impact, by definition, but can bring about the enrichment of the atmosphere with water through the above-mentioned chemical reaction.
The impact energy of planetesimals suffices to melt, at lease, the impact sites of the protoplanet surface (i.e., magma ponds). For planetesimals larger than $\sim$10~km, because of the self-blanketing effect by ejected materials from the impact craters, the energy deposited by planetesimals accumulates in the interior \citep[][]{Safronov78,Kaula79}. 
Detailed modelling demonstrates that magma ponds (or partial magma oceans) exist on accreting sub-Mars-mass protoplanets 
\citep[][]{Senshu+02}. 
When it comes to close-in planets around M stars, recent formation models show that those planets likely undergo giant collisions of lunar-size embryos during periods of accretion and migration \citep[e.g.][]{Ogihara+Ida2009,Tian+Ida2015}. 
Such giant collisions would make the planetary surfaces entirely molten (i.e., global magma oceans) \citep[][]{Tonks+93}. 
Thus, a rocky protoplanet likely has molten areas of its surface and thereby produces water through oxidation of atmospheric hydrogen.

\subsection{Composition uniformity}
\label{sec:discussion_uniformity}
If enrichment occurs only in the deep atmosphere and vertical mixing is inefficient, our assumption of composition uniformity is invalid and the effect of enrichment is limited.
Especially, as for water production via the interaction between the atmosphere and magma ocean, inefficient mixing leads to low production efficiency because little fresh hydrogen is supplied to the deep atmosphere.
Nevertheless, vigorous convection and thereby mixing occur up to a certain altitude during a phase of rapid planetesimal accretion (i.e., high luminosity).
According to Fig.~\ref{fig:Ad_model1}, for example, it is suggested that a Mars-mass solid protoplanet has an entirely convective atmosphere even for $\XHHO$ = 0.01, provided planetesimal accretion rate is high enough ($L$ = $1 \times 10^{27}$~erg/s). 
As planetesimal accretion rate declines, the upper atmosphere becomes radiative and the boundary between the upper radiative and lower convective regions (or the tropopause) goes down to an altitude with the temperature of 400--500~K. 
As shown in Fig.~\ref{fig:xH2O_Menv_Thp}(a), if the tropopause is located at such an altitude, the atmosphere is less massive by 1--2 orders of magnitude than the completely mixed atmosphere. 
In contrast, for more massive protoplanets of 0.3~$M_\oplus$, the atmospheric mass is still large even for $T_\mathrm{hp}$ = 400--500~K (see Fig.~\ref{fig:xH2O_Menv_pseudo}b).

Condensation has been shown to have a great impact on the atmospheric mass (see Sect.~\ref{sec:parameter study}). 
However, when water is produced only at the bottom of the atmosphere, it has to be transported upwards. 
Then the temperature becomes low enough;
most of the vapour would condense at that height and little vapour may be transported further up (i.e., cold trap).
In this case, the moist-adiabatic region is significantly narrowed and the atmospheric mass may be reduced by an order of magnitude or more. 
Thus, since the behaviour of water in the moist-convective region critically affects the mass of the captured water, a detailed investigation is of great importance.

Mixing in the magma ocean is also an important factor because it affects the efficiency of oxygen supply to the surface.
During the planetesimal accretion stage, the magma ocean is considered to be strongly convective due to its low viscosity and the large amount of deposited accretion energy \citep[e.g., see][]{Solomatov2007}.
However, an increase in surface temperature due to the water production would lead to stratifying the upper layer of the magma ocean, which may limit oxygen supply and thus water production.

In summary,
both for the atmosphere and magma ocean, mixing processes are the key to understanding how much water terrestrial planets can finally acquire from the disc gas. 
In the upper atmosphere, accreting gas flows and collisions of planetesimals or embryos cause some turbulence, which may lead to stirring the atmosphere. Also, as for the magma ocean, impacts of accreting planetesimals and sedimentation of metal droplets formed at the surface would stir such a stratified layer. 
Detailed treatments of these processes should be future work. 

\subsection{Precipitation of condensates}
\label{sec:discussion_precipitation}
Whether or not precipitation occurs depends on the settling velocity of droplets and the convective velocity. 
In the case of the structure of the highly enriched atmosphere ($\XHHO=0.75$) presented in Fig.~\ref{fig:PT_model2}(b), for example,
the moist-convective region extends downwards from ($P$, $T$) $\sim$ (10~$\si{Pa}$, 200~K) to ($1 \times 10^5$~{Pa}, 300~K); the corresponding radial distances from the protoplanet centre are $\sim 10R_\oplus$ and $\sim 2R_\oplus$, respectively.
From these values, the settling velocity of droplets is estimated to be on the order of  $\sim$ 10--100~cm/s even for mm-size droplets.
Note that we have used the same method described in \cite{Ormel2014} for this estimate, assuming the density of droplets is $1~\si{g/cm^3}$.

The convective velocity, on the other hand, can be estimated from the mixing length theory as~\citep[e.g.,][]{Kippenhahn+Weigert1990} 
\begin{equation}
    v_\text{conv} = 
        \qty[ \frac{\alpha \varphi}{4}\frac{P}{\rho c_P T}\frac{F_\text{conv}}{\rho}]^{1/3},
\end{equation}
where $\alpha$ is the ratio of pressure scale height to mixing length, $\varphi=-(\pdv*{\ln \rho}{\ln T})_P$, $c_P$  is the specific heat at constant pressure, and $F_\text{conv}$ is the convective energy flux.
Given that both $\alpha$ and $\varphi$ are on the order of unity, $c_P$ is on the order of $10^7\si{erg/g.K}$ %assuming the proper mixture
for ideal mixtures of hydrogen and water vapour and $F_\text{conv}=L/(4\pi r^2)$, 
the convective velocity would be $v_\text{conv}\sim 10^2$--$10^3$~m/s in the moist-convective region in the case of $L=\SI{1e27}{erg/s}$. 
It turns out that the convective velocity is much higher than the sedimentation velocity of droplets.

Thus, the condensed water is more likely to remain in the upper atmosphere than to precipitate in the case of such high energy fluxes.
In the low energy flux case, where the upper layer of the atmosphere is radiative, the droplets would settle down to the inner dry-adiabatic region.
However, since the settling velocity is quite small as estimated above,
it would take more than 10~Myr for the droplets to sink through such an extended atmosphere.
Therefore, even in the low energy flux case, 
the droplets are likely to remain in the atmosphere during the protoplanetary disc lifetime~\citep[a few Myr; ][]{Mamajek2009}.

\subsection{Atmospheric escape}
\begin{figure*}
    \includegraphics[width=2\columnwidth]{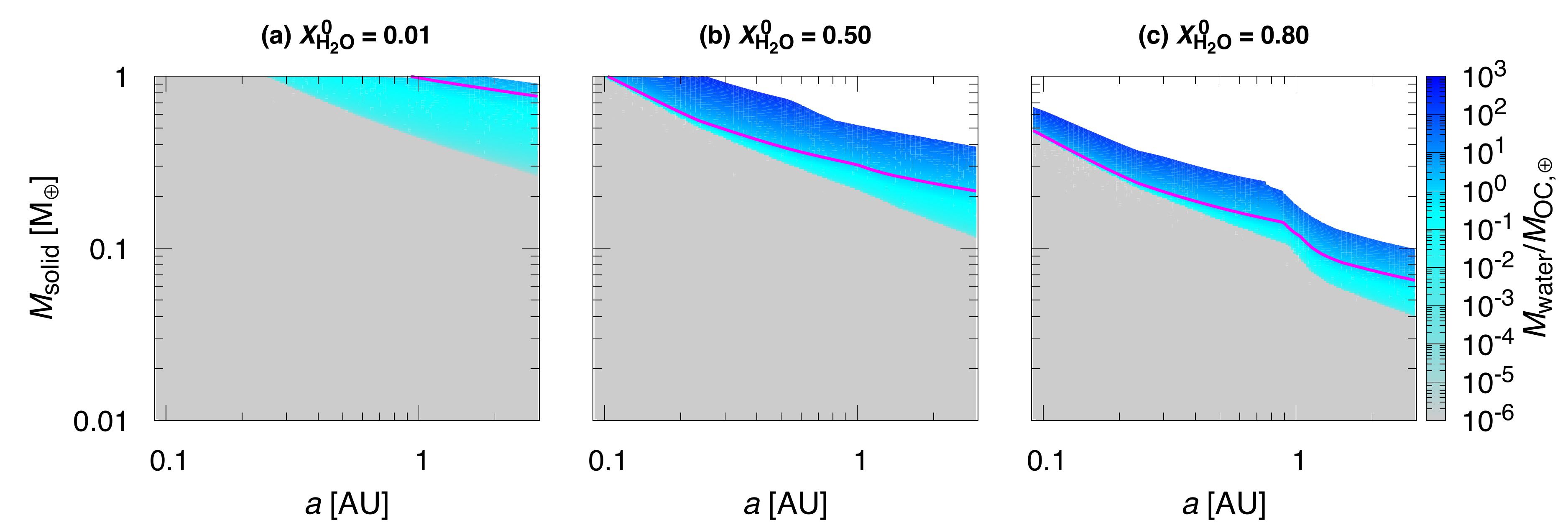}
    \caption{Same as Fig.~\ref{fig:Mw_before_loss}, but for $f_g = 1 \times 10^{-8}$ and with the effect of atmospheric escape due to stellar irradiation.}
    \label{fig:Mw_after_loss}
\end{figure*}   
% \begin{figure}
%     \centering
%     \includegraphics[width=\columnwidth]{MwMap_pseudo_lost.pdf}
%     \caption{Same as Fig.~\ref{fig:Mw_after_loss}(c), but for the case including the condensed water precipitation.}
%     \label{fig:Mw_pseudo}
% \end{figure}

Once the disc gas dissipates, planets are subjected to high energy irradiation such as X-ray and UV from host stars, which drives atmospheric escape (or photo-evaporation). 
Here we assess the impact of the photo-evaporation on the final mass of the captured water 
%\kimura{
in a simple way. Detailed treatment of the photo-evaporation process is beyond the scope of this study;
we also neglect other escape processes for simplicity \citep[e.g., see][]{Loyd+2020}.
%}

First, we calculate the atmospheric mass for $f_g=10^{-8}$, 
which corresponds to the XUV optical thickness at 0.1~AU (measured from the central star) of  unity.
Then we estimate the atmospheric escape rate due to XUV irradiation from the host star by adopting a simple energy-limited escape model~\citep{Sekiya+1980,Watson+1981};
\begin{equation}
    \dMesc
        = \frac{\varepsilon\Lxuv \Rc^3}{4G\Mc a^2}.
       \label{eq:dMesc}
\end{equation}
Here we have assumed that the absorption of stellar XUV occurs at an altitude that is much smaller than the planetary radius, $\Rc$ \citep[e.g.][]{Sekiya+1980,Watson+1981}.
We set the stellar XUV luminosity $\Lxuv$ to be $1 \times 10^{-3}$ times the stellar bolometric luminosity $L_\ast$, following the saturated XUV flux~\citep{Pizzolato+2003,Scalo+2007,Jackson+2012} and $\varepsilon = 0.1$. 
%\kimura{
Note that we adopt 0.1 for $\varepsilon$ because such a value is often used in the literature \citep[e.g.,][]{Tian+Ida2015,Hori+Ogihara2020} and supported by numerical simulations, at least, for atmospheres of Earth and super-Earths \citep[e.g.,][]{Owen+Alvarez2016,Bolmont+2017}; however, detailed investigation remains to be done for sub-Earths.
%}
The time evolution of $L_*$ is taken from the table given by \cite{Baraffe+1998}.
We integrate Eq.~\eqref{eq:dMesc} from 5~Myr to 1~Gyr, assuming that the disc gas prevents stellar XUV radiation from reaching the planet before 5~Myr, and, then, subtract the amount of the lost gas from the atmospheric mass calculated above.

The mass of water that survives the atmospheric escape is shown as a function of $\Msolid$ and $a$ for three different choices of $\XHHO$ in Fig.~\ref{fig:Mw_after_loss}.
Although the range of planetary mass with which the planet can hold a significant amount of water is narrower compared to Fig.~\ref{fig:Mw_before_loss}, it turns out that even sub-Mars-mass planets can still keep water comparable in amount with the Earth's oceans in the case of $\XHHO$ = 0.8. 

%\kimura{
We should note here that as mentioned in Sect.~\ref{Sec:Method}, the disc temperature at the timing of disk dispersal (typically $\sim$ 5~Myr) could be lower than that given by Eq.~\eqref{eq:TB}, according to recent theoretical studies suggesting lower stellar luminosity in the pre-MS phase~\citep[e.g.,][]{Kunitomo+2017}.
Lower stellar luminosities result in more massive primordial atmospheres because of the cooler nebula gas lower outer boundary temperature $T_\mathrm{out}$, and probably in smaller escape rate because of weak UV emission from the less active star.
Thus, the survived mass of water could be larger.
Furthermore,
%}
we have assumed here that water is always in the vapour form. In reality, however, as the atmosphere cools, water condenses and rains down to the surface (i.e., ocean formation), surviving the atmospheric escape. Also, the planet in the blank area of Fig.~\ref{fig:Mw_after_loss} acquires much more atmospheric gas including water via the runaway accretion. 
With such effects, less massive planets would be able to retain 1-$M_{\rm OC, \oplus}$ water.

\subsection{Ingassing}
Volatiles such as water and hydrogen are known to dissolve well in molten silicate (or magma). Thus, there occurs ingassing of the captured disc gas and the produced water into the magma ocean~\citep{Sharp2017,Olson+Sharp2019}.
Recently \cite{Olson+Sharp2019} developed the atmospheric ingassing-outgassing model for protoplanets that have a primordial atmosphere and suggested that the amount of ingassed hydrogen and water could be as much as several Earth ocean masses.
The ingassed water may avoid the escape process caused by stellar irradiation from the central star.
Moreover, the ingassing causes additional disc gas inflow which may result in further production of water.
Thus the water content of terrestrial planets can be larger than estimated in the previous sections, provided if both the water production and ingassing occur effectively.

In contrast, if the ingassing occurs more quickly than vertical mixing in the atmosphere, the produced water is hardly transported upwards and significant atmosphere enrichment hardly occurs.
In this case, it is difficult for the surface temperature to be kept high enough and the water production mechanism never works well.
The exchange of water between the atmosphere and the interior should also be investigated in the future.

\section{Conclusion} \label{Sec:Conclusioin}

The primordial atmosphere of the nebular origin of rocky planets is not always hydrogen-dominated. 
It would likely be highly enriched with water through oxidation of the atmospheric hydrogen with oxidising rocky materials from incoming planetesimals or the magma ocean. 
Thermodynamically normal oxygen buffers are known to produce water comparable in mass to or more than hydrogen~\citep{Sasaki1990,Ikoma+Genda2006}.
In this study, we have simulated the 1D structure of such an enriched atmosphere and investigated the effects of water enrichment on the atmospheric properties, in particular the amounts of water, of rocky protoplanets. %\kimura{
%Especially, 
We have %focused on 
supposed
sub-Earth-mass planets around M dwarfs of $0.3M_\odot$.
%}

For the atmosphere uniformly enriched with water, we have found the followings:
\begin{enumerate}
    \item The atmospheric mass increases by more than one order of magnitude, as the water mass fraction in the atmospheric gas ($\XHHO$) increases from 0.01 to 0.8. This means that the amount of captured water increases by more than two orders of magnitude.
    \item 
    The mass of captured water increases more significantly with $\XHHO$ in relatively cool circumstellar regions because of water condensation in the upper atmosphere.
    \item Even a Mars-mass planet can obtain water comparable in mass to the Earth's oceans if $\XHHO \gtrsim 0.6$. 
    \item In the classic habitable zone ($\sim$ 0.1-0.2~AU), even for planets of 0.3-0.5~$M_\oplus$, the captured water
    %the atmosphere with captured water 
    survives
    the atmospheric escape processes due to disc gas dispersal and stellar UV irradiation.
\end{enumerate}

The above results suggest the possibility that there are more %terrestrial planets 
sub-Earth-mass planets with Earth-like water contents in extrasolar systems than previously predicted. 
In particular, M dwarfs may be able to harbour habitable planets, although sub-Earths are expected to be a majority around M dwarfs~\citep{Raymond+2007,Ida+2005,Alibert+Benz2017,Miguel+2019}. 
Since it is predicted that capture of icy planetesimals  is unlikely to bring moderate amounts of water to planets in the habitable zone around M dwarfs, this water production process would have a great importance for habitable planet formation around such stars.
%\kimura{
However, our results are based on the assumptions of efficient water production and efficient material mixing in the atmosphere and magma ocean.
Also, we have neglected the effects of water vapour ingassing.
%}
How much water a terrestrial planet really obtains
and retains against atmospheric loss % \remove{, however,}
depends on these factors and also on photoevaporation efficiency,
%\remove{water production efficiency and material mixing in the atmosphere and magma ocean}
detailed investigation for which % detailed investigation
should be made in the future.

\section*{Acknowledgements}
We thank the anonymous referee for her/his careful reading and constructive comments that helped improve the manuscript.
This work is supported by JSPS KAKENHI Nos.~JP18H05439 and JP18H01265 and JSPS Core-to-core Program `International Network of Planetary Sciences.' 
T.~K. is supported by International Graduate Program for Excellence in Earth-Space Science (IGPEES).

\section*{Data availability}
The data underlying this article will be shared on reasonable request to the corresponding author.
%%%%%%%%%%%%%%%%%%%%%%%%%%%%%%%%%%%%%%%%%%%%%%%%%%

%%%%%%%%%%%%%%%%%%%% REFERENCES %%%%%%%%%%%%%%%%%%

% The best way to enter references is to use BibTeX:

\bibliographystyle{mnras}
\bibliography{refer} % if your bibtex file is called example.bib

% Alternatively you could enter them by hand, like this:
% This method is tedious and prone to error if you have lots of references
% \begin{thebibliography}{99}
% \bibitem[\protect\citeauthoryear{Author}{2012}]{Author2012}
% Author A.~N., 2013, Journal of Improbable Astronomy, 1, 1
% \bibitem[\protect\citeauthoryear{Others}{2013}]{Others2013}
% Others S., 2012, Journal of Interesting Stuff, 17, 198
% \end{thebibliography}

%%%%%%%%%%%%%%%%%%%%%%%%%%%%%%%%%%%%%%%%%%%%%%%%%%

%%%%%%%%%%%%%%%%% APPENDICES %%%%%%%%%%%%%%%%%%%%%

\appendix
\section{Dust Enhancement Factor}\label{sec:appendix}
Here we give an outline of the method presented in \cite{Ormel2014} to derive the dust enhancement/depletion factor $f$ that we have used in Eq.~\eqref{eq:kappa}.

The basic equation for the radial distribution of dust grains' characteristic mass $m^*$ is given by Eq.~\eqref{eq:grain-mass}.
In this study we assume that the monomer density and size are 3~g cm$^{-3}$ and 1~$\mu$m, respectively.
It is assumed that the settling velocity $v_\mathrm{settl}$ is equal to the terminal velocity and the dust growth occurs due to the Brownian motion and the differential drift motion \citep[see][]{Ormel2014}. 
To estimate $\dot{M}_{\text{dep}}$, 
we assume that planetesimals disintegrate in a narrow region
in the atmosphere and the mass flux profile is given by
\begin{equation}
    \dv{\dot{M}_{\text{dep}}}{r} = -\rhogas \dot{M}_{\text{plts}}
        P_{\ln}(\sigma ; \sigma_{\text{crit}},\delta),
\end{equation}
where $\dot{M}_{\text{plts}}$ is the planetesimal accretion rate, $\sigma$ is the column density of atmospheric gas, $\sigma_{\text{crit}}$ is the column density beyond which planetesimal disintegration occurs, 
and $\delta$ is a parameter controlling the width of this function.
The distribution function $P_{\ln}$ is 
\begin{equation}
    P_{\ln}(x; \beta, \delta)
    = \frac{1}{\delta x \sqrt{2\pi}} 
        \exp[ -\frac{1}{2\delta^2} \left\{\ln (\frac{x}{\beta})\right\}^2].
\end{equation}
We set $\sigma_{\text{crit}} = 100~\si{g/cm^2}$ and $\delta = 0.2$, following the standard parameter set used in \cite{Ormel2014}.
Although they choose relatively small value of $\sigma_\text{crit}$ to see the effect of grain mass deposition, 
%\kimura{
which corresponds to the case that planetesimals break up at relatively upper regions of the atmosphere,
%}
we use the same value for simplicity. This hardly affects our results.

With $m^*$ and $v_{\text{settl}}$ obtained, we calculate the spatial mass density of dust grains by
\begin{equation}
    \rho_{\text{gr}} = \frac{\dot{M}_{\text{dep}}(r)}
                            {4\pi r^2 v_{\text{settl}}}.
\end{equation}
Then we obtain the dust-to-gas ratio $Z_{\text{gr}}=\rho_{\text{gr}}/\rhogas$, and $f = Z_{\text{gr}}/Z_{\text{gr0}}$, where $Z_{\text{gr0}}$ is the dust-to-gas ratio of the disk gas.
We set $Z_{\text{gr0}} = 0.01$ in this study.

%%%%%%%%%%%%%%%%%%%%%%%%%%%%%%%%%%%%%%%%%%%%%%%%%%

% Don't change these lines
\bsp	% typesetting comment
\label{lastpage}
\end{document}